\documentclass[12pt,a4wide]{iopart}
\usepackage[dvips]{graphicx}
\usepackage{pst-plot}
\usepackage{bm}
\usepackage{mathrsfs}
\usepackage{amsfonts}
\usepackage{amssymb}
\usepackage{lscape}
\usepackage{epic,eepic}
\usepackage{hyperref}
\usepackage{bezier}
\usepackage{pstricks}
\usepackage{dcolumn}

\newcommand{\ket}[1]{|#1\rangle}
\newcommand{\bra}[1]{\langle#1|}
\newcommand{\braket}[1]{\langle#1\rangle}

\begin{document}

\title[Many-body interactions and nuclear structure]{Many-body interactions and nuclear structure}

\author{M.~Hjorth-Jensen} 
\address{Department of Physics, University of Oslo, P.O.~Box 1048
Blindern, N-0316 Oslo, Norway}
\address{Center of Mathematics for Applications, University of Oslo,
P.O.~Box 1056 Blindern, N-0316 Oslo, Norway}

\author{D.J.~Dean}
\address{Physics Division, Oak Ridge National Laboratory, P.O. Box 2008, 
Oak Ridge, Tennessee 37831, USA}  

\author{G.~Hagen}
\address{Physics Division, Oak Ridge National Laboratory, P.O. Box 2008, 
Oak Ridge, Tennessee 37831, USA}  

\author{S.~Kvaal} 
\address{Center of Mathematics for Applications, University of Oslo,
P.O.~Box 1056 Blindern, N-0316 Oslo, Norway}


\begin{abstract}
This article presents several challenges to  
nuclear many-body theory and our understanding of the stability of nuclear matter. 
In order to achieve this, we present five different 
cases, starting with an idealized toy model.
These cases
expose problems that need to be understood
in order to match recent advances in nuclear theory with
current experimental programs in low-energy nuclear physics.

In particular, we focus on our current understanding, or lack thereof, 
of many-body forces, and how they evolve as functions of the number of particles.
We provide 
examples of discrepancies between theory and experiment and 
outline some selected perspectives for future research directions.

\end{abstract}

\date{\today}
\maketitle

\section{Introduction}\label{sec:introduction}

A central issue in  basic nuclear physics research 
is to understand the limits of stability of matter starting from its basic 
building blocks, either represented by effective degrees of freedom such as various
hadrons or employing  the underlying theory of the strong interaction, namely quantum chromodynamics. 
To achieve this implies the development of a comprehensive description 
of all nuclei and their reactions, based on a strong interplay between theory and experiment. 
This interplay should match the research conducted at
present and planned experimental facilities, where one of the  aims is to  
study unstable and rare isotopes. These nuclei can convey crucial
information about the stability of nuclear matter, 
but are difficult to produce experimentally since they can have extremely
short lifetimes. They exhibit also unsual neutron-to-proton ratios that are very different from their
stable counterparts.
Furthermore, these
rare nuclei lie at the heart of nucleosynthesis processes in 
the universe and are therefore an important component in the 
puzzle of matter generation in the universe. 

We do expect that these
facilities will
offer unprecendeted data on weakly bound systems and the limits of stability.
To interpret such a  wealth of experimental data and point 
to new experiments that can shed light on various
properties of matter requires a reliable and predictive theory. If a theoretical model is capable
of explaining a wealth of experimental data, one can thereafter analyze the results in terms
of specific components of, say, the nuclear forces and extract simple physics pictures
from complicated many-body systems.

To better understand  the rationale for this article, it is important to keep in mind 
the particularity of nuclear physics. 
Basic nuclear physics research, as conducted today, is 
very diverse 
in nature, with experimental facilities which include accelerators, 
reactors, and underground laboratories.  
This diversity 
reflects simply the rather
complex nature of the nuclear forces among protons
and neutrons. These generate a broad range and variety
in the nuclear phenomena that we observe, from energy scales of
several gigaelectronvolts (GeV) to a 
few kiloeletronvolts (keV). Nuclear physics is thus a classic example of what we would call multiscale physics.
The many scales  
pose therefore a severe challenge to nuclear many-body theory and what are dubbed 
{\em ab initio} descriptions~\footnote{With {\em ab initio} we do mean methods which
allow us to solve exactly or within controlled approximations, 
either the non-relativistic Schr\"odinger's equation or the relativistic Dirac
equation for many interacting particles. The input to these methods is a given Hamiltonian 
and relevant degrees of freedom such as neutrons and protons and various mesons.} of nuclear systems.
Examples of key physics issues which need to be addressed by nuclear theory are:
\begin{itemize}
\item How do we derive the in-medium nucleon-nucleon interaction from basic principles?
\item How does the nuclear force depend on the proton-to-neutron ratio?
\item What are the limits for the existence of nuclei?
\item How can collective phenomena be explained from individual motion?
\item Can we understand shape transitions in nuclei?
\end{itemize}

In order to deal with the above-mentioned problems from an {\em ab initio} standpoint, it is our firm belief
that nuclear many-body theory needs to meet some specific criteria in order to be credible.
Back in 2004 two of the present authors co-authored  a preface to a collection of articles
from a workshop on many-body theories, see Ref.~\cite{preface2004} for more details. Seven specific requirements to theory were presented. These were:
\begin{itemize}
\item
A many-body theory should be fully microscopic and start with present two- and three-body
interactions derived from, {\it e.g.}, effective field theory;
\item The theory can be improved upon systematically, e.g., by inclusion of
three-body interactions and more complicated correlations;
\item It allows for a description of both closed-shell 
systems and valence systems;
\item For nuclear systems where shell-model studies are the only feasible ones,
viz., a small model space requiring an effective interaction, 
one should be able to
derive  effective two- and three-body 
equations and interactions for the nuclear shell
model;
\item It is amenable to parallel computing;
\item It can be used to generate excited spectra for nuclei like 
where many shells are involved. (It is hard for the traditional shell model
to go beyond one major shell.  The inclusion of several shells may imply 
the need of  complex effective interactions
needed in studies of weakly bound systems); and
\item Finally, nuclear structure results should be used in marrying microscopic 
many-body results with reaction studies. This will be another hot topic
of future {\it ab initio} research.
\end{itemize}
 Six years have elapsed since these requirements were presented. Recent advances in nuclear theory have made redundant most of the points listed above. 
They are actually included in many calculations, either fully or partly.  On the other hand, in the last five years, we have witnessed considerable progress in nuclear theory. Of relevance to this article
are the developments of effective field theories \cite{epelbaum2009a,epelbaum2009b}, lattice quantum chromodynamics calculations, with the possibility to constrain specific parameters of the nuclear forces \cite{ishii2007,ishii2009,kaplan2009} and of  many-body theories applied to nuclei,
see for example Refs.~\cite{bogner2010,navratil2009,hagen2009,barbieri2009}. There are also several efforts to link standard {\em ab initio} methods with density functional theories, or more precisely, energy density functional theories, see Refs.~\cite{jacek2009,thomas2009,witek2009}.

These developments have led us to formulate some key intellectual issues and further requirements we feel nuclear theory needs
to address. The key issues are:
\begin{enumerate}
\item Is it possible  to link lattice quantum chromodynamics calculations with   
effective field theories in order to better understand the nuclear forces to be used in a many-body theory?
\item A nuclear force derived from effective field theory is normally constructed with a
specific cutoff in energy or momentum space. Most interactions have a cutoff $\Lambda$ in the range 
$\Lambda\sim 500$-$600$ MeV. The question we pose here is how well do we understand 
the link between this cutoff and a specific Hilbert space (the so-called model space) 
used in a many-body calculation?
\item The last question is crucially linked with our understanding of many-body forces and leads us 
to the next question. Do we understand how many-body forces evolve as we add more and more particles?
Irrespective of whether our Hamiltonian contains say two- and three-body interactions, a truncated
Hilbert space results in missing many-body correlations. To understand how these correlations 
evolve as a function of the number of particles is crucial in order to
provide a predictive many-body theory.
\item Finally, in order to deal with systems beyond closed-shell nuclei with or without some few valence nucleons, we need to link {\em ab initio} methods with density functional theory. 
\end{enumerate}
With respect to requirements, we find it timely to request that a proper many-body theory 
should provide uncertainty
quantifications. For most methods, this means to provide an estimate of the error due to the truncation made in the single-particle basis and the truncation made in limiting
the number of possible excitations. The first point, as shown by Kvaal \cite{kvaal2009}, can be rigorously proven for a chosen single-particle basis. The second point is more difficult and is normally justified
{\em a posteriori}. As an example, in  coupled-cluster calculations, a truncation is made in terms
of various particle-hole excitations. Only specific sub-clusters of excitations are included to infinite order, see for example Ref.~\cite{bartlett2007}. Whether this subset of excitations is sufficient or not can only be justified after the calculations have been performed.

This article does not aim at answering the above questions. 
Rather, our goal is to raise the awareness about these issues since it is our belief that they can lead to a more predictive nuclear many-body theory. 
Our focus is on the third topic, and we illustrate the problems which can arise 
via a simple toy model in Section \ref{sec:toymodel}. Thereafter, we discuss four possible
physics cases where the effect of missing many-body forces can be studied 
theoretically and benchmarked through existing and planned experiments.  The physics cases are all linked
with studies of isotopic chains of nuclei, with several closed-shell nuclei accessible to {\em ab initio} calculations. Our physics cases are discussed in Section \ref{sec:physicscases}. We start 
with the chain of oxygen isotopes, from $^{16}$O to
$^{28}$O. The next cases deal with isotopes in the $pf$ shell, namely the Ni isotopes ($^{48}$Ni, $^{56}$Ni, $^{68}$Ni, and $^{78}$Ni) and the Ca isotopes ($^{40}$Ca, $^{48}$Ca, $^{52}$Ca, $^{54}$Ca, and $^{60}$Ca). We conclude with
the chain of tin isotopes from $^{100}$Sn to $^{140}$Sn. For isotopes like $^{28}$O, $^{54}$Ca, $^{60}$Ca, and
$^{140}$Sn, data on binding energies are missing. We try to give here a motivation why one should attempt
to measure these nuclei.

Our conclusions and perspectives are presented in the last section. 

\section{A simple toy model}\label{sec:toymodel}
We start with a model described by a simple two-body  Hamiltonian.  This model catches the basic features concerning 
the connection between many-body forces and the size of the model space. In particular we will stress
the link between a given effective Hilbert space and missing many-body effects.  
This section, although it forms the largest part of this article, conveys the problems which can arise 
in many-body calculations with truncated spaces and effective interactions.   

The model we present mimicks the effects seen in 
standard shell-model calculations with an effective interaction, either with  a valence model space \cite{hko1995} or a
so-called no-core model space \cite{navratil2009}. 

\subsection{Hamiltonian}

The Hamiltonian acting in the complete Hilbert space (usually infinite
dimensional) consists of an unperturbed one-body part, $\hat{H}_0$,
and a perturbation $\hat{V}$. The goal is to obtain an effective
interaction $\hat{V}_{\mathrm{eff}}$ acting in a chosen model or
valence space. By construction, it gives a set of eigenvalues
which are identical to (a subset of) those of the complete problem.

If we limit ourselves to, at most, two-body interactions, our Hamiltonian  is 
then represented by the following operators

\[
\hat{H} = \sum_{\alpha\beta}\langle \alpha |h_0|\beta\rangle a_{\alpha}^{\dagger}a_{\beta} +\frac{1}{4}\sum_{\alpha\beta\gamma\delta}\langle \alpha\beta| V|\gamma\delta\rangle a_{\alpha}^{\dagger}a_{\beta}^{\dagger}a_{\delta}a_{\gamma},
\]
where $a_{\alpha}^{\dagger}$ and $a_{\alpha}$, etc.~are standard fermion creation and annihilation operators, respectively,
and $\alpha\beta\gamma\delta$ represent all possible single-particle quantum numbers. 
The full single-particle space is defined by the completeness relation
$\hat{{\bf 1}} = \sum_{\alpha =1}^{\infty}|\alpha \rangle \langle \alpha|$.
In our calculations  we will let  the single-particle states $|\alpha\rangle$
be eigenfunctions of  the one-particle operator $\hat{h}_0$.

The above Hamiltonian 
acts in turn on various many-body Slater determinants constructed from the single-basis defined by the one-body
operator $\hat{h}_0$.    
We can then diagonalize a  two-body problem in  a large space and project  
out via a similarity transformation
an effective two-body interaction.  The interaction 
acts in a much smaller set of single-particle states. 
The two-particle model space $\mathcal{P}$ is defined by an operator 
\[
\hat{P} =   \sum_{\alpha\beta =1}^{m}|\alpha\beta \rangle \langle \alpha\beta|,
\]
where we assume that $m=\dim(\mathcal{P})$ and the full space is defined by
\[
\hat{P}+\hat{Q}=\hat{{\bf 1}}.
\]

\begin{figure*}[htbp]
\vspace{1.0cm}
 \setlength{\unitlength}{1cm}
 \begin{picture}(15,14)
 \thicklines
\put(-0.6,1){\makebox(0,0){$p=1$}}
\put(-0.6,2){\makebox(0,0){$p=2$}}
\put(-0.6,3){\makebox(0,0){$p=3$}}
\put(-0.6,4){\makebox(0,0){$p=4$}}
\put(-0.6,5){\makebox(0,0){$p=5$}}
\put(-0.6,6){\makebox(0,0){$p=6$}}
\put(-0.6,7){\makebox(0,0){$p=7$}}
\put(-0.6,8){\makebox(0,0){$p=8$}}
\put(-0.6,9){\makebox(0,0){$p=9$}}
\put(-0.6,10){\makebox(0,0){$p=10$}}
\put(-0.6,11){\makebox(0,0){$p=\dots$}}
\put(16,8){\makebox(0,0){$\mathcal{Q}$}}
\put(16,3){\makebox(0,0){$\mathcal{P}$}}
\put(0.8,1){\circle*{0.3}}
\put(0.8,2){\circle*{0.3}}
\put(1.7,1){\circle*{0.3}}
\put(1.7,2){\circle*{0.3}}
\put(5.0,1){\circle*{0.3}}
\put(5.9,1){\circle*{0.3}}
\put(5.0,4){\circle*{0.3}}
\put(5.0,2){\circle{0.3}}
\put(5.9,3){\circle*{0.3}}
\put(5.9,2){\circle{0.3}}
\put(9.2,1){\circle*{0.3}}
\put(10.1,3){\circle*{0.3}}
\put(9.2,4){\circle*{0.3}}
\put(10.1,8){\circle*{0.3}}
\put(10.1,1){\circle{0.3}}
\put(9.2,2){\circle{0.3}}
\put(10.1,2){\circle{0.3}}
\put(13.4,7){\circle*{0.3}}
\put(14.3,6){\circle*{0.3}}
\put(13.4,9){\circle*{0.3}}
\put(14.3,9){\circle*{0.3}}
\put(13.4,1){\circle{0.3}}
\put(14.3,1){\circle{0.3}}
\put(13.4,2){\circle{0.3}}
\put(14.3,2){\circle{0.3}}
\dashline[+1]{2.5}(0,1)(15,1)
\dashline[+1]{2.5}(0,2)(15,2)
\dashline[+1]{2.5}(0,3)(15,3)
\dashline[+1]{2.5}(0,4)(15,4)
\dashline[+1]{2.5}(0,5)(15,5)
\dashline[+1]{2.5}(0,6)(15,6)
\dashline[+1]{2.5}(0,7)(15,7)
\dashline[+1]{2.5}(0,8)(15,8)
\dashline[+1]{2.5}(0,9)(15,9)
\dashline[+1]{2.5}(0,10)(15,10)
\thinlines
\dashline{0.1}(0,5.5)(15,5.5)
\dashline{0.1}(0,11)(15,11)
 \end{picture}
\caption{Schematic plot of the possible single-particle levels with double degeneracy.
The filled circles indicate occupied particle states while the empty circles 
represent vacant particle(hole) states.
The spacing between each level $p$ is constant in this picture. 
The first five single-particle levels define a possible model space, indicated by the label $\mathcal{P}$.
The remaining states span the excluded space $\mathcal{Q}$. 
The first state to the left represents
a possible ground state representation for a four-fermion system. In the second state to the left,
one pair is broken by the interaction. 
All single-particle orbits belong to the model space. The two remaining four-particle states represents single-particle excitations to the excluded space, either by breaking two pairs or breaking one pair and exciting one pair (rightmost state). \label{fig:schematic}}
\end{figure*}
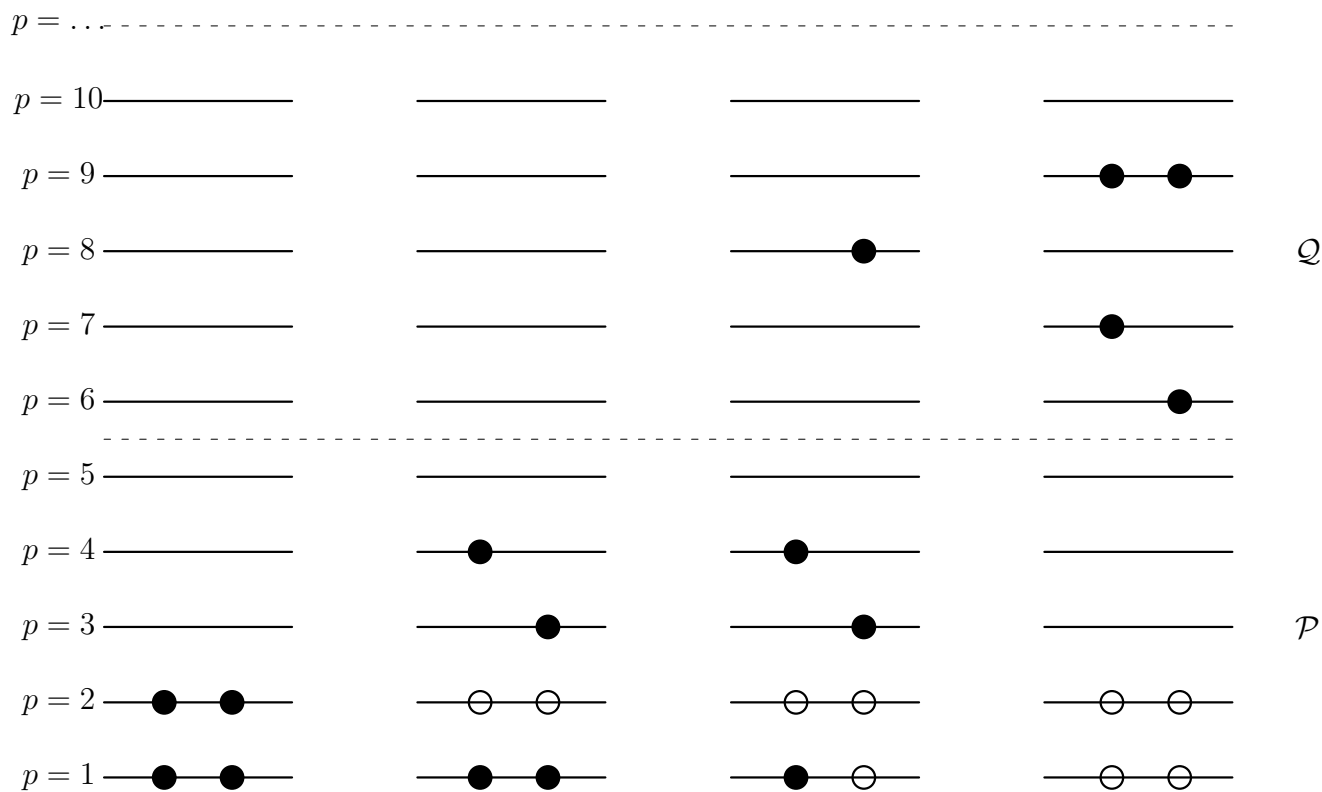

Our specific model consists of $N$ doubly degenerate and equally spaced
single-particle levels labelled by $p=1,2,\dots$ and spin $\sigma=\pm
1$.  These states are schematically portrayed in
Fig.~\ref{fig:schematic}.  The first five single-particle levels
define a possible model space indicated by the label $\mathcal{P}$.  

We write
the Hamiltonian as 
\[ \hat{H} = \hat{H}_0 + \hat{V} , \]
where
\[
\hat{H}_0=\xi\sum_{p\sigma}(p-1)a_{p\sigma}^{\dagger}a_{p\sigma}
\]
and 
\[
\hat{V}=-\frac{1}{2}g\sum_{pq}a^{\dagger}_{p+}
a^{\dagger}_{p-}a_{q-}a_{p+} -\frac{1}{2}
f\sum_{pqr}\left(a^{\dagger}_{p+}a_{p-}^{\dagger}a_{q-}a_{r+} + \mathrm{h.c.}\right).
\]
Here, $H_0$ is the unperturbed Hamiltonian with a spacing between
successive single-particle states given by $\xi$, which we may set to
a constant value $\xi=1$ without loss of generality. The two-body
operator $\hat{V}$ has two terms. The first term represents the
pairing contribution and carries a constant strength $g$. 
(It easy to extend our model to include a state dependent interaction.)  The indices
$\sigma=\pm$ represent the two possible spin values. The first term of
the interaction can only couple pairs and excites therefore only two
particles at the time, as indicated by the rightmost four-particle
state in Fig.~\ref{fig:schematic}. There, a pair is excited to the
state with $p=9$.  The second interaction term, carrying a constant
strength $f$, acts between a set of particles with opposite spins and
allows for the breaking of a pair or just to excite a single-particle
state.  The spin of a given single-particle state is not changed.
This interaction can be interpreted as a particle-hole interaction if
we label single-particle states within the model space as
hole-states. The single-particle states outside the model space are
then particle states.

In our model we have kept both the interaction strength and the single-particle level as constants.
In a realistic system like a nucleus, this is not the case; however, if a harmonic oscillator basis
is used, as done in the no-core shell-model calculations \cite{navratil2009}, at least the single-particle
basis mimicks the input to realistic calculations.  

\subsection{Effective Hamiltonians}

We now consider a general $A$-body situation, of which our model is
just an example, where a Hilbert space of finite dimension $n$ is
given along with an $A$-body Hamiltonian $\hat{H}$ with spectral
decomposition given by
\[ \hat{H} = \sum_{k=1}^n E_k \ket{\psi_k}\bra{\psi_k}, \] where
$\{\ket{\psi_k}\}_{k=1}^n$ is an orthonormal basis of eigenvectors and
where $E_k$ are the corresponding eigenvalues. We choose the dimension
$n$ to be finite for simplicity, but the theory may be generalized to
infinite dimensional settings where $\hat{H}$ has a purely discrete
spectrum. 

The Hilbert space is divided into the model space $\mathcal{P}$ and
its complement, denoted by $\mathcal{Q}$.  
We assume again that $m=\dim(\mathcal{P})$.

The effective Hamiltonian $\hat{H}_{\mathrm{eff}}$ is defined in the $\mathcal{P}$-space
\emph{only}, and by definition its eigenvalues are identical to the $m$ eigenvalues 
of $\hat{H}$. This is equivalent to $\hat{H}_{\mathrm{eff}}$ being given by
\begin{eqnarray*} 
  \hat{H}_{\mathrm{eff}} &:=& \hat{P}\mathcal{H}\hat{P} \\
  &=& \hat{P} e^{-\hat{G}} \hat{H} e^{\hat{G}} \hat{P}, 
\end{eqnarray*}
where $\mathcal{H}$ is assumed to obey the de-coupling equation
\[ 
 \hat{Q}\mathcal{H}\hat{P} = 0. 
\] 
If the latter is
satisfied, the $\mathcal{P}$-space is easily seen to be invariant under
$\mathcal{H}$, and since similarity transformations preserve
eigenvalues, $\hat{H}_{\mathrm{eff}}$ is seen to have $m$ eigenvalues of $\hat{H}$. 

Without loss of generality, we assume that the eigenvalues $E_k$ of
$\hat{H}$ are arranged so that $\hat{H}_{\mathrm{eff}}$, which is non-Hermitian in
general, has the spectral decomposition
\[ \hat{H}_{\mathrm{eff}} = \sum_{k=1}^m E_k \ket{\phi_k}\tilde{\bra{\phi_k}}, \] where
$\{\ket{\phi_k}\}_{k=1}^m$ is a basis for the $\mathcal{P}$-space, and
where $\braket{\phi_k|\tilde{\phi_\ell}} = \delta_{k,\ell}$ defines
the bi-orthogonal basis $\{\ket{\tilde{\phi}_k}\}_{k=1}^m$.

The similarity transform operator $\exp(\hat{G})$ is, of course, not
unique; $E_k$, $k=1,\cdots,m$ can be chosen in many ways, and even if the
effective eigenvector $\ket{\phi_k}$ is chosen to be related
to $\ket{\psi_k}$, there is still great freedom of choice left.

Assume that we have determined the eigenvalues $E_k$, $k=1,\ldots,m$
that $\hat{H}_{\mathrm{eff}}$ should have. Two choices of the corresponding
$\ket{\phi_k}$ are common: The Bloch-Brandow choice, and the canonical
Van Vleck choice, resulting in ``the non-Hermitian'' and ``the
Hermitian'' effective Hamiltonians, respectively. For a discussion of these approaches see 
Refs.~\cite{vanvleck1929,bloch1958a,bloch1958b,brandow1967,brandow1977,brandow1979,klein1974,shavitt1980}.

In the Bloch-Brandow scheme, the effective eigenvectors are simply
chosen as
\[ \ket{\phi_k} := \hat{P}\ket{\psi_k}, \] which gives meaning
whenever $\hat{P}\ket{\psi_k}$ defines a basis for
$\mathcal{P}$-space. In this case, $\hat{G} = \hat{\omega}$, where
$\hat{\omega} = \hat{Q}\hat{\omega} \hat{P}$, defined by
\[ 
\hat{\omega} \hat{P}\ket{\psi_k} := \hat{Q}\ket{\psi_k}, \quad k =
1,\cdots,m. 
\]

In contrast, the canonical Van Vleck effective Hamiltonian chooses a
certain orthogonalization of $\{\hat{P}\ket{\psi_k}\}_{k=1}^m$ as
effective eigenvectors \cite{kvaal2008b}. In this case, $\hat{G} =
\mathrm{arctanh}(\hat{\omega}-\hat{\omega}^\dag)$, which relates the two
effective Hamiltonians to each other. The canonical effective interaction $\hat{H}_{\mathrm{eff}}$ minimizes
the quantity $\Delta$ defined by
\begin{equation}
  \Delta(\ket{\chi_1}, \cdots, \ket{\chi_m}) := \sum_{k=1}^m \|
\ket{\chi_k} - \ket{\psi_k} \|^2, 
\label{eq:delta}
\end{equation}
where the minimum is taken with
respect to all orthonormal sets of $\mathcal{P}$-space
vectors $\chi$. In fact, it can be taken as the definition \cite{kvaal2008b}. The
Bloch-Brandow effective eigenvectors, on the other hand, yield the
\emph{global} minimum of $\Delta$.

We have not yet specified \emph{which} of the eigenvalues of $H$ is to
be reproduced by $\hat{H}_{\mathrm{eff}}$. In general, we would like it to reproduce
the ground state and the other lowest eigenstates of $H$ if $m>1$. We
consider this question in some detail in the following sections.

But first, we define the effective \emph{interaction} $\hat{V}_{\mathrm{eff}}$ as
\[ \hat{V}_{\mathrm{eff}} := \hat{H}_{\mathrm{eff}} - \hat{P}\hat{H}_0\hat{P},\] where
$[\hat{H}_0,\hat{P}]=0$ is assumed. This is satisfied whenever the
model space is spanned by Slater determinants being eigenvectors of
$\hat{H}_0$, such as in the model studied in this article.

Finding $\hat{H}_{\mathrm{eff}}$ is equivalent to solving the original problem. In order to be
useful, we need some sort of approximation scheme to find $\hat{H}_{\mathrm{eff}}$. It
is common to use many-body perturbation techniques such as folded
diagrams \cite{hko1995}, but as these suffer from convergence problems,
the no-core shell model community \cite{navratil2009}
has developed the so-called
sub-cluster approximation to $\hat{V}_{\mathrm{eff}}$. This is a non-perturbative
approach, and utilizes the many-body nature of $\hat{H}$. The
resulting effective interaction is often referred to as the Lee-Suzuki
effective interaction \cite{leesuzuki}.

In our model, $\mathcal{P}$ is defined by restricting the allowed
levels accessible for the $A$ particles under study. Thus,
\[ \mathcal{P}^{(A)} := \left\{
  \ket{(p_1,\sigma_1)\cdots(p_A,\sigma_A) } \; : \; p_k \leq N_P
  \right\}, \]
where $N_P \leq N$ is the number of levels accessible in the model
space. This is the way model spaces in general are built up, simply restricting
the single-particle orbitals accessible. It is by no means the only
possible choice. On the other hand, this way of defining the model
space has a very intuitive appeal, as it naturally leads to a view of
$\hat{H}_{\mathrm{eff}}$ as a renormalization of $\hat{H}$. It also gives a natural
relation between model spaces for different $A$, which is absolutely
necessary for the sub-cluster effective Hamiltonian to be meaningful.

The effective Hamiltonian is seen to be an $A$-body
operator in general, even though $H$ itself may contain only two-body
operators. Thus, $\hat{V}_{\mathrm{eff}}$ can be written in its most general form as
\[ \hat{V}^{(A)}_{\mathrm{eff}} = \sum_{\alpha_1,\cdots,\alpha_A}
\sum_{\beta_1,\cdots,\beta_A} u^{\alpha_1,\cdots}_{\beta_1,\cdots}
a^\dag_{\alpha_1} \cdots a^\dag_{\alpha_A} a_{\beta_A} \cdots
a_{\beta_1}, \] where $\alpha_k=(p_k,\sigma_k)$ and $u^{\alpha_1,\cdots}_{\beta_1,\cdots}$
represent a specific matrix element. 
The approximation idea is then to obtain instead an $a$-body effective
interaction $\hat{V}_{\mathrm{eff}}^{(a)}$, where $a<A$, and view this as an
approximation to $\hat{V}_{\mathrm{eff}}^{(A)}$. This leads to
\[ \hat{V}^{(A)}_{\mathrm{eff}} \approx
\frac{\left(\begin{array}{c}A\\2\end{array}\right)}{\left(\begin{array}{c}A\\a\end{array}\right)\left(\begin{array}{c}a\\2\end{array}\right)}\hat{V}^{(a)}_{\mathrm{eff}},  \]
which is a much simpler operator, usually obtainable exactly by
large-scale diagonalization of the $a$-body Hamiltonian.

The remaining question is which eigenpairs of $H^{(a)}$ should be
reproduced by $\hat{H}_{\mathrm{eff}}^{(a)}$, and which approximate eigenvectors should
be used. There is no unique answer to this. The ``best'' answer would be
for each problem to require a complete knowledge of the conserved
observables of the many-body Hamiltonian \cite{kvaal2009,kvaal2008b}. 

On the other hand, if $\hat{V}$ is a small perturbation, that is, we let
$\hat{V} \mapsto \lambda \hat{V}$ and consider an adiabatic turning on
by slowly increasing $\lambda$, then it is natural to choose the
eigenvalues developing adiabatically from $\lambda=0$. Indeed,
$\hat{V}_{\mathrm{eff}}^{(a)}$ is then seen to be identical to a class of $a$-body
terms in the perturbation series for the full $\hat{V}_{\mathrm{eff}}^{(A)}$ to
infinite order \cite{klein1974,shavitt1980}. The problem is, there is no way in
general to decide \emph{which} eigenvalues have developed
adiabatically from $\lambda=0$, and we must resort to a heuristic
procedure.

Two alternatives present themselves as obvious candidates: Selecting
the smallest eigenvalues, and selecting the eigenvalues whose
eigenvectors have the largest overlap $\braket{\psi_k|P|\psi_k}$ with
$\mathcal{P}$. Both are equivalent for sufficiently small $\lambda$,
but the eigenvalues will cross in the presence of so-called intruder
states for larger $\lambda$
\cite{sw1972,sw1973,schaefer1974}. Moreover, the presence of perhaps
unknown constants of motion will make the selection by eigenvalue
problematic, as exact crossings may lead us to select eigenpairs with
$\hat{P}\ket{\psi_k}=0$, which makes $\hat{H}_{\mathrm{eff}}^{(a)}$ ill-defined. We
therefore consider selection by model space overlap to be more robust
in general. These considerations are, however, not important for our main conclusions.
 
The non-Hermitian Bloch-Brandow effective Hamiltonian runs into problems in the sub-cluster
approach since the interpretation of $\hat{V}_{\mathrm{eff}}^{(a)}$ as an interaction
requires hermiticity when applied to an $A$-body problem -- otherwise,
eigenvalues will be complex unless $a=A$.

See Ref.~\cite{kvaal2008b} for details on the algorithm for computing
the effective interactions.

\subsection{Pure pairing interaction}

We first set $f\equiv 0$, giving a pure pairing Hamiltonian. We study
an $A=5$ body problem with $N=8$ levels and a model space consisting
of the $N_P=5$ first levels. We compute $\hat{V}_{\mathrm{eff}}^{(a)}$ for $a\leq 4$
and compare their properties. We study here only the selection of eigenvectors using the overlap
scheme.

\begin{figure}
\includegraphics[width=0.5\textwidth]{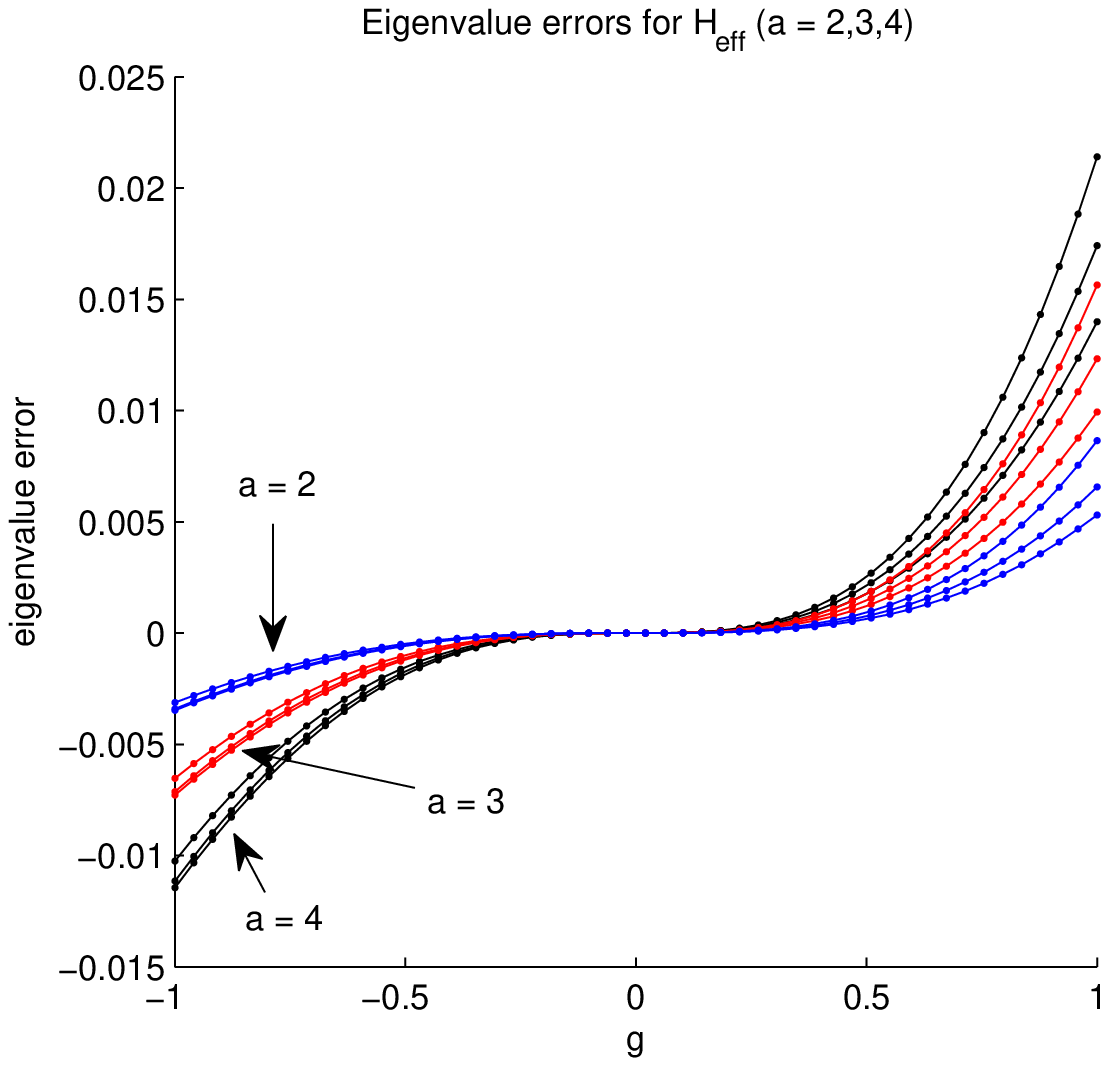}%
\includegraphics[width=0.5\textwidth]{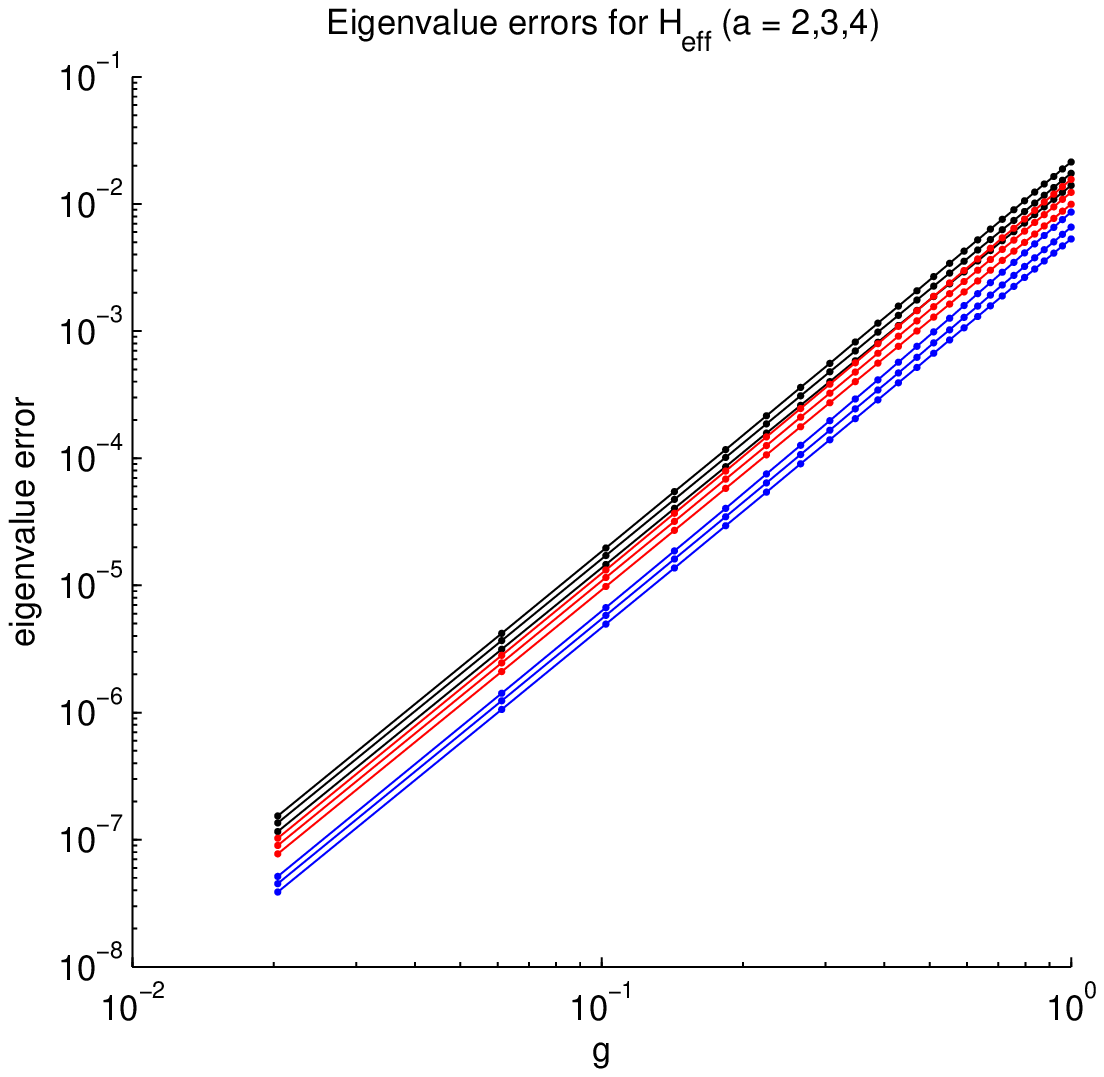}
\caption{(Color online) Left: Error in the 3 lowest eigenvalues using
  $\hat{V}_{\mathrm{eff}}^{(a)}$, $a=2$ (black), $a=3$ (red), and $a=4$ (blue) compared
  to exact eigenvalues ($A=5$). Right: logarithmic plot for
  $g>0$, showing almost perfect $g^3$-behavior.\label{fig:errors1}}
\end{figure}

In Fig.~\ref{fig:errors1} the error in the three first eigenvalues for
each $a$ are shown as a function of $g\in [-1,1]$. For $g>0$, a
double-logarithmic plot reveals an almost perfect $g^3$-behaviour of
all errors. Larger values of $a$ give smaller errors, as one would expect, but
only by a constant factor. Thus, all the $\hat{V}_{\mathrm{eff}}^{(a)}$ seem to be
equivalent to perturbation theory to second order in the strength $g$ with respect to
accuracy. This order is constant, even though the complexity of
calculating $\hat{V}_{\mathrm{eff}}^{(a)}$ increases by orders of magnitude.  Most of the physical correlations are thus
well represented by a two-body effective interaction. This is expected since a pairing-type
interaction favors strong two-particle clusters. The choice of a constant pairing strength enhances
also these types of correlations. Three-body and four-body clusters tend to be small. 

By investigating the remaining eigenvalues, we confirm that \emph{all}
effective eigenvalues behave in the described way, with only small
variations.
\begin{figure}
\includegraphics[width=0.5\textwidth]{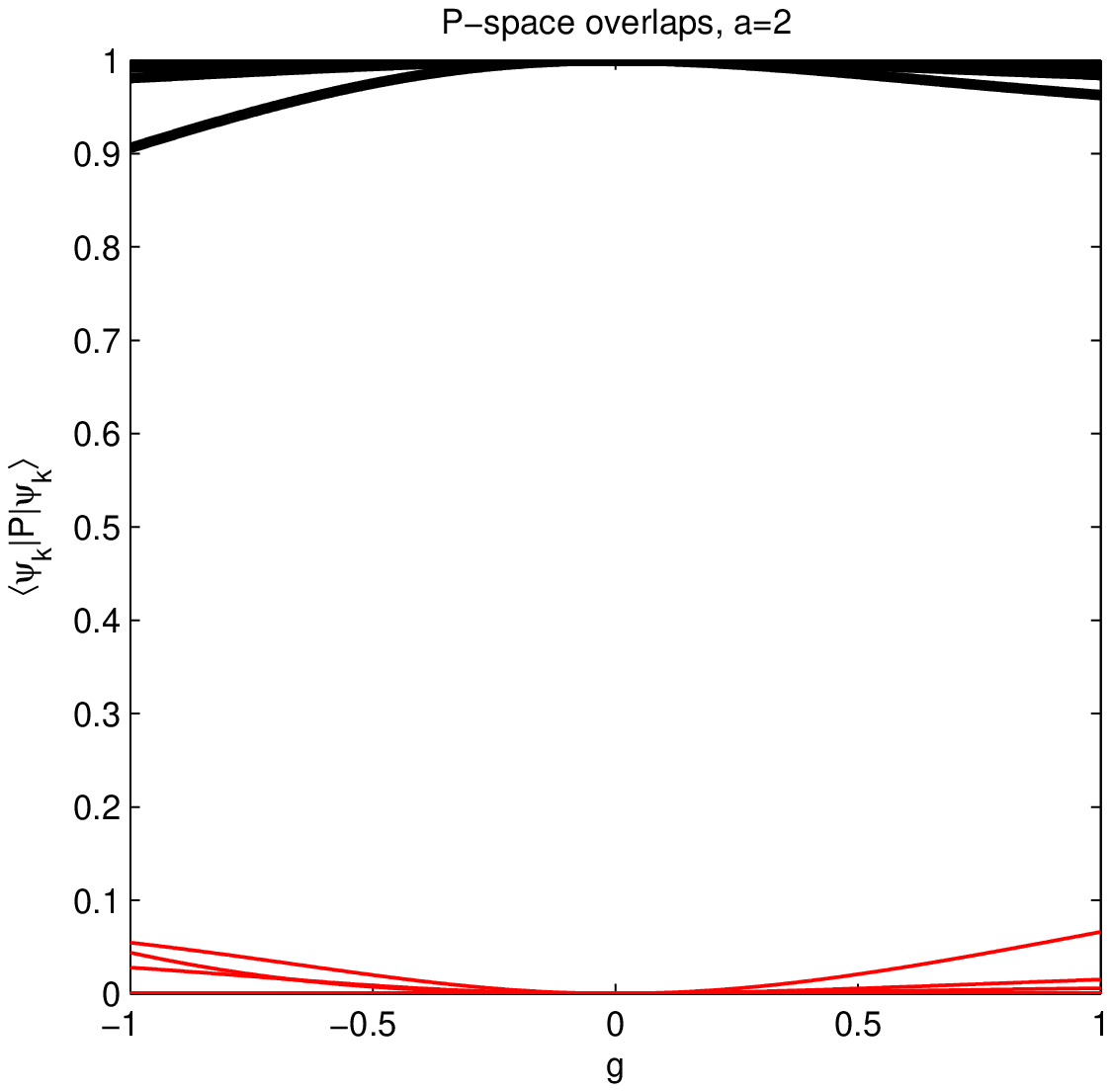}%
\includegraphics[width=0.5\textwidth]{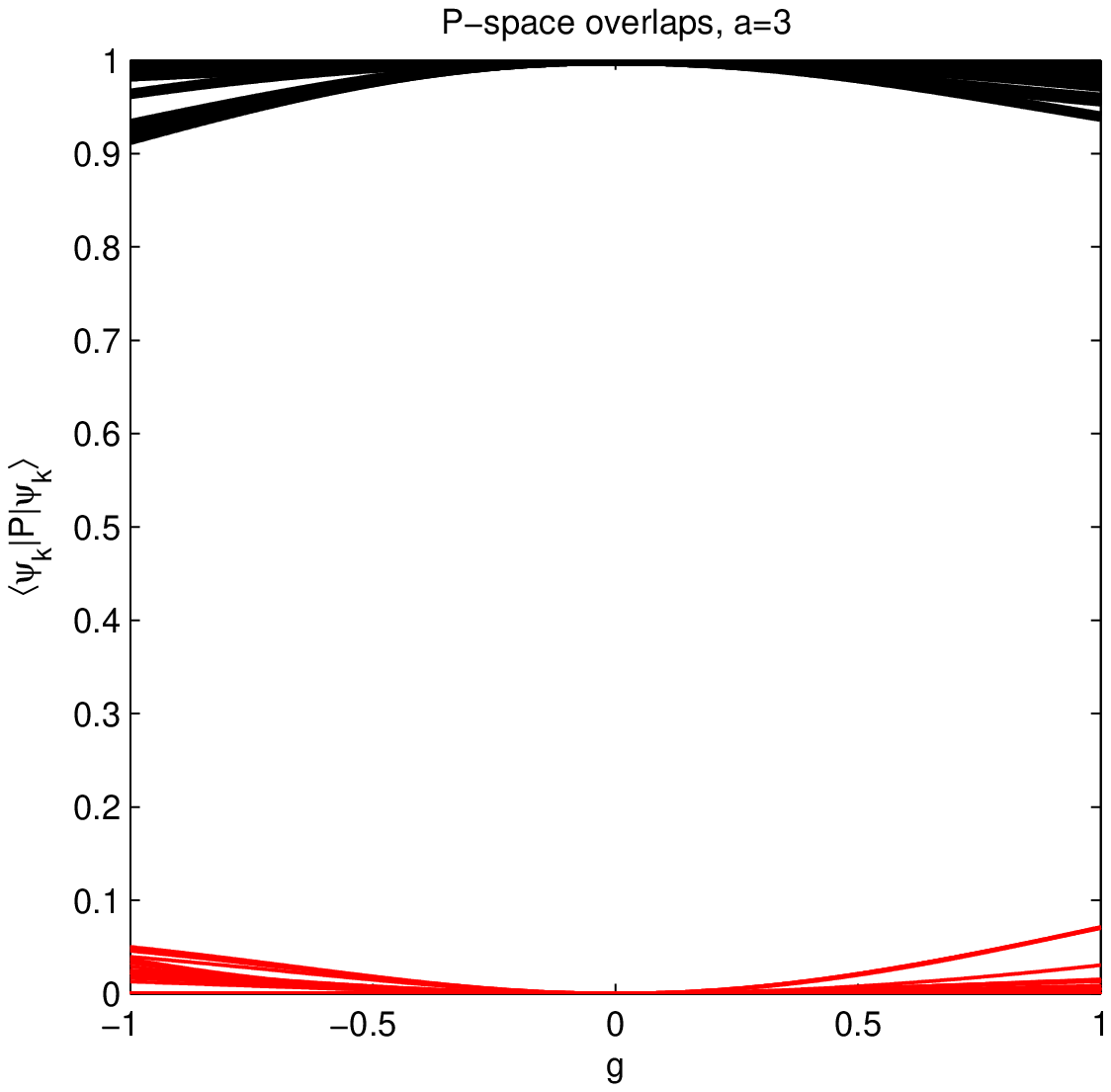}
\includegraphics[width=0.5\textwidth]{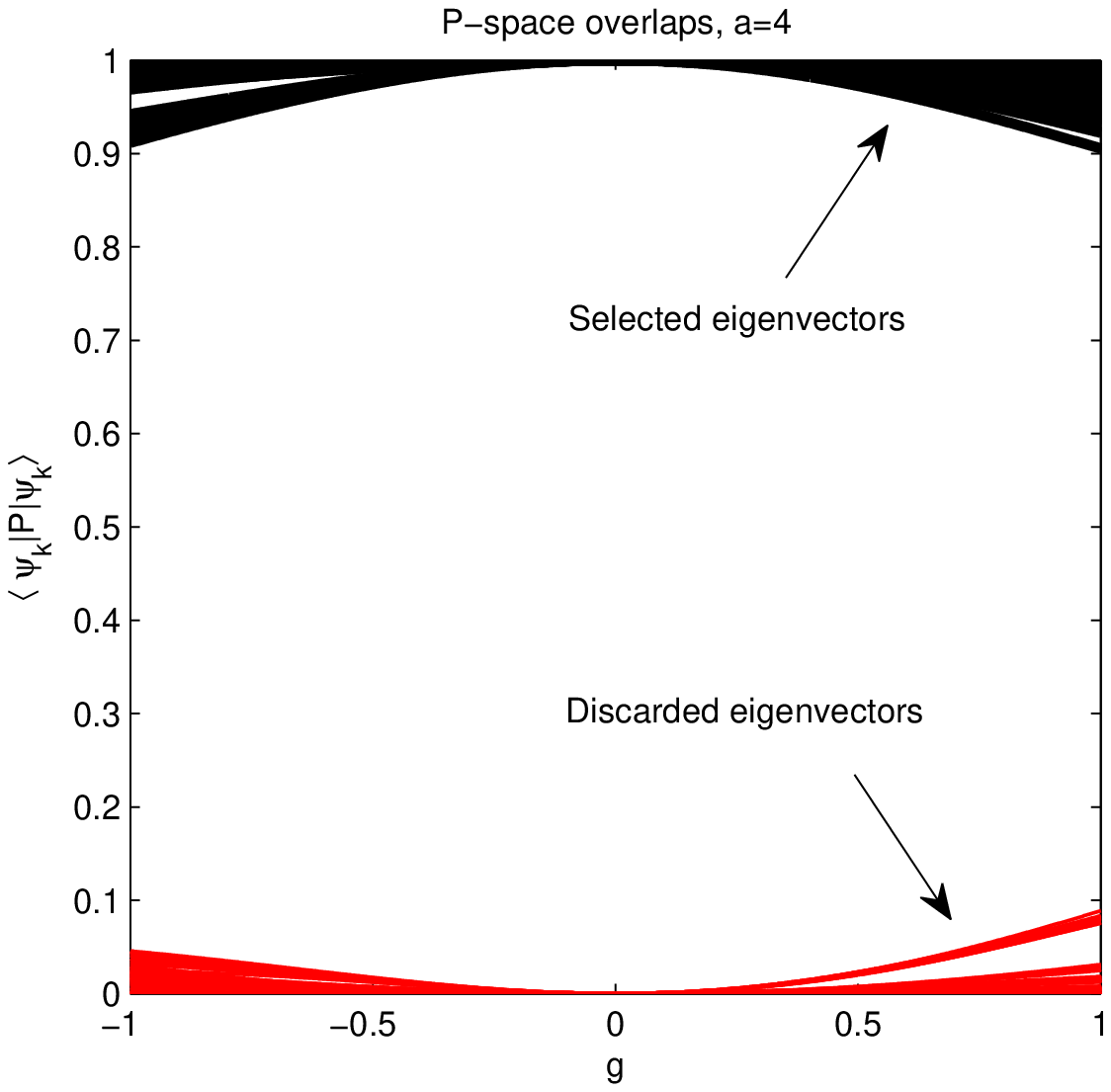}%
\includegraphics[width=0.5\textwidth]{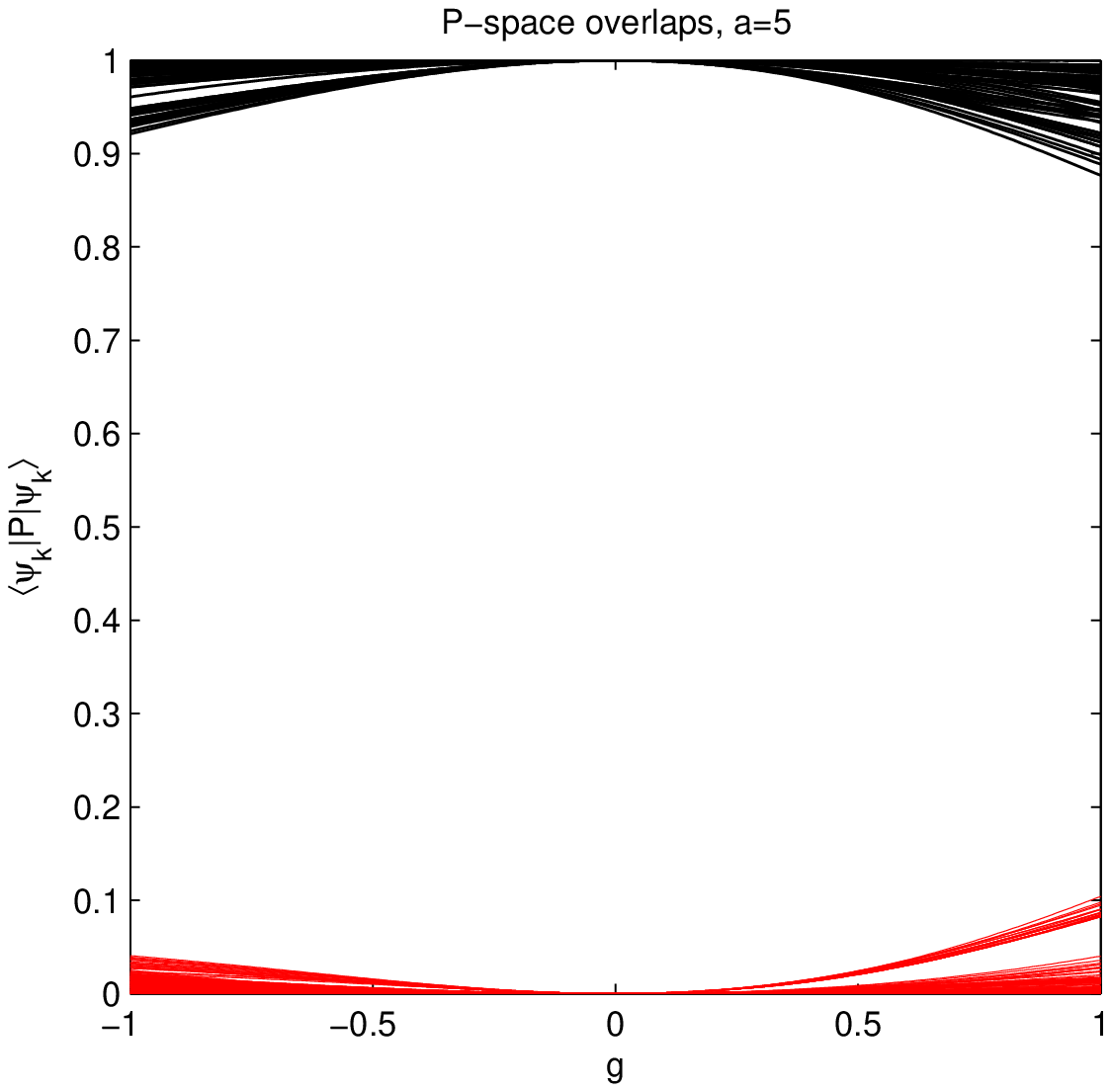}
\caption{(Color online) Model space overlaps for $a\leq 5$,
  $f=0$. Black lines are selected model space vectors' overlaps $\braket{\psi_k|P|\psi_k}$, 
  while red lines
  are excluded space overlaps. Each line corresponds to one
  eigenvector of the $a$-body Hamiltonian. The $\mathcal{P}$-space
  selections and $\mathcal{Q}$-space selections are clearly separated
  from each other.\label{fig:overlaps1}}
\end{figure}
To assess the quality of the overlap selection scheme in this case, we
graph the overlaps $\braket{\psi_k|P|\psi_k}$ 
of the selected and discarded eigenvectors
for each $a\leq 5$ as a function of $g$ in
Fig.~\ref{fig:overlaps1}. Here, the selected overlaps are shown in black,
while the discarded are colored red. We show only overlaps in the
total spin $S=0$ ($a=2,4$) and $S=1/2$ ($a=3,5$) channels for
clarity. The selections for the  $\mathcal{P}$-space and the $\mathcal{Q}$-space are
well separated for all $g$, meaning that the selection scheme manages
to follow the eigenpairs adiabatically from $g=0$, where
$\braket{\psi_k|P|\psi_k}$ is either zero or unity. Moreover, the view of the
$\mathcal{P}$-space as being ``effective'' is sensible.
However, for very large values of $g$, the graphs of the selected and discarded
eigenvectors cross, meaning that the effective interaction view is broken. 

We conclude that the effective interaction $\hat{V}_{\mathrm{eff}}^{(a)}$ works very
well in the pure pairing Hamiltonian case, as long as the strength $g$
is sufficiently small. However, there is not much to gain with respect
to accuracy by going to $a>2$ compared to $a=2$.  That is, a two-body effective interaction
captures the relevant physics when only a pairing force is involved. 
In nuclear physics, the pairing interaction is rather strong, and the typical relation between
the single-particle spacing $\xi$ around the Fermi surface and the strength of the  nuclear pairing 
interaction $g$  
is roughly $|\xi/g|\sim 1-5$. In light nuclei like $^{16}$O, the shell-gap is $11.25$ MeV, and the average
$p$-shell and $sd$-shell effective interactions are 
of the order of $1-5$ MeV in absolute value \cite{khj2008}.  
In the region of the rare earth nuclei, typical single-particle spacings are of the order of some few hundred keV. The interaction matrix elements
are of the same size in absolute value.  
Choosing thereby a parameter $g$ slightly larger or smaller than one captures the essential
physics produced by a pairing interaction around the  Fermi surface in nuclei. 
 
\subsection{Pairing and particle-hole interaction}

\begin{figure}
\includegraphics[width=0.5\textwidth]{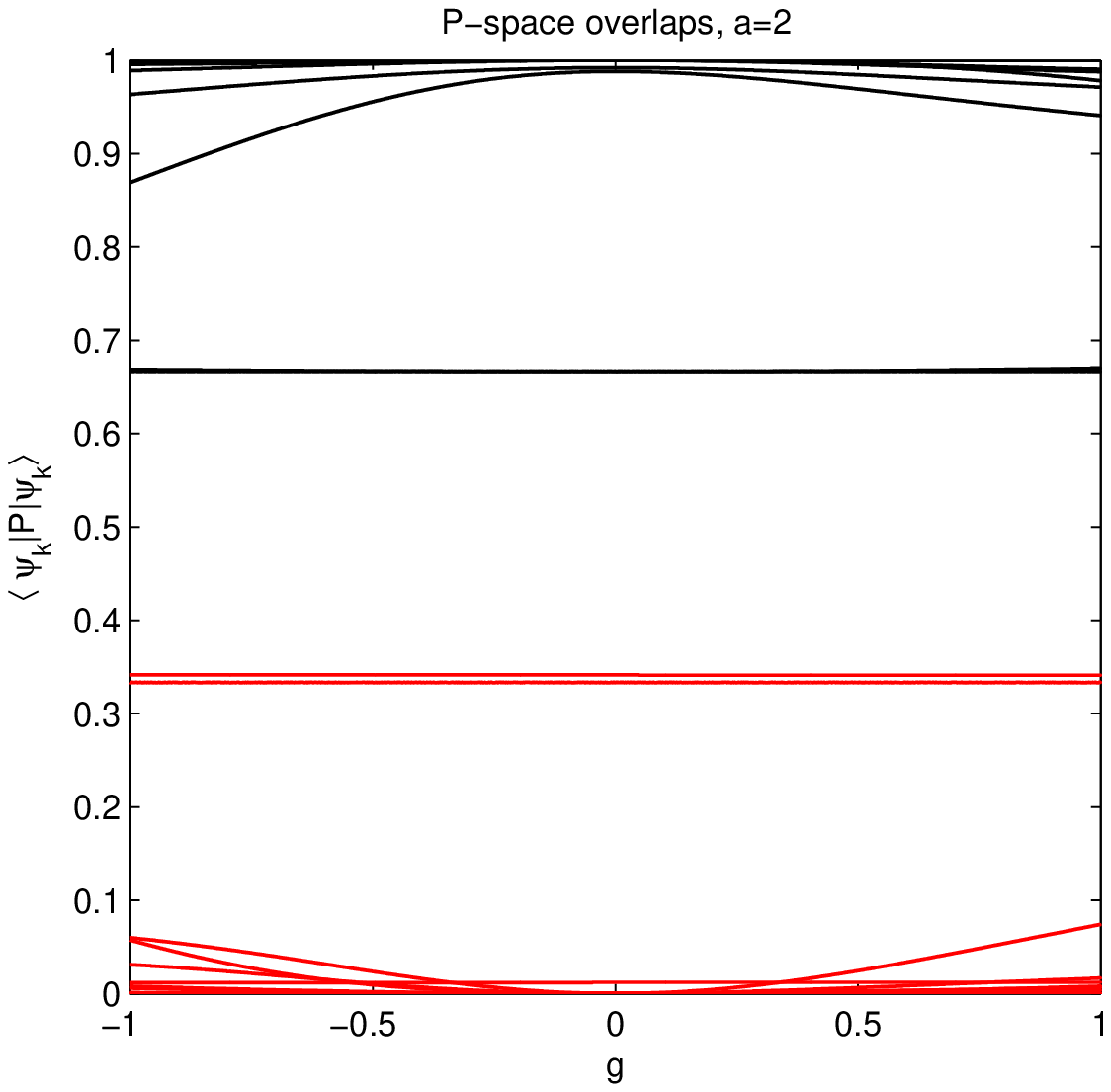}%
\includegraphics[width=0.5\textwidth]{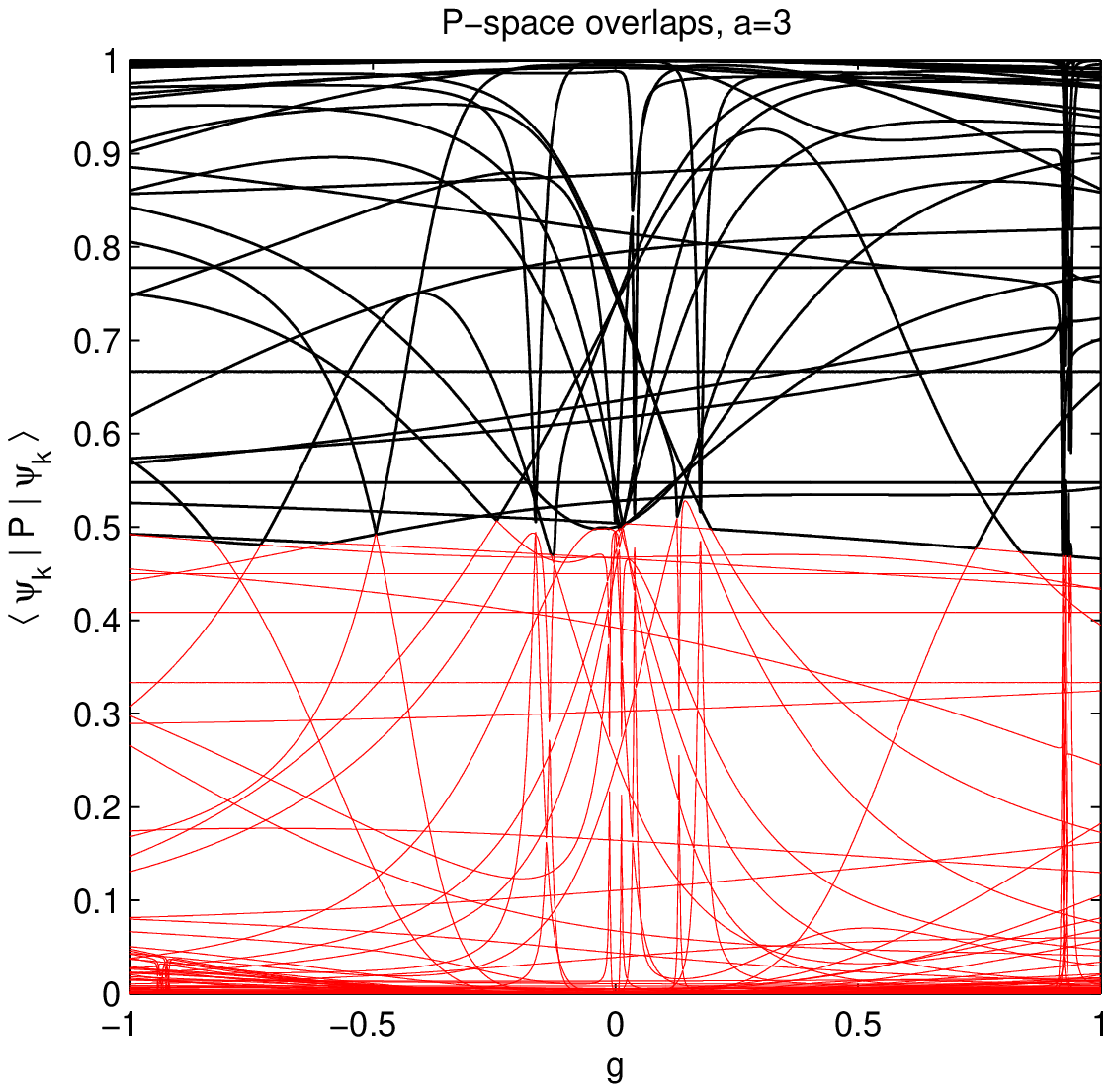}
\includegraphics[width=0.5\textwidth]{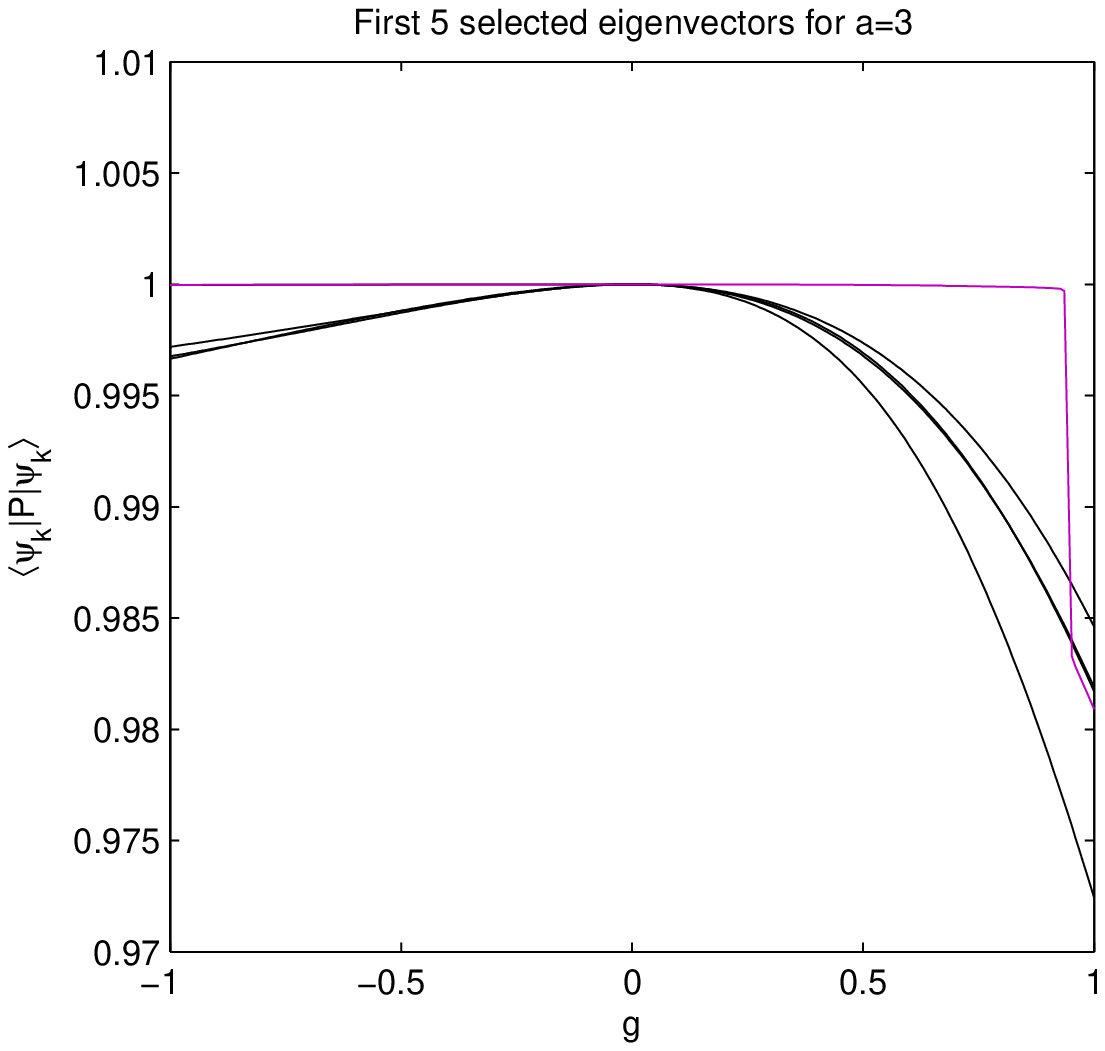}
\caption{(Color online) Model space overlaps for $a=2,3$,
  $\alpha=0.05$. Top figures: Black lines are selected model space
  overlaps, while red lines are excluded space overlaps. The
  $\alpha=0$ picture (cf.~top row of Fig.~\ref{fig:overlaps1}) is
  radically changed. For $a=4$ (not shown) the picture is even more
  complicated. These graphs show that it is difficult to choose a
  proper model space when the particle-hole interaction is turned on.
  Bottom figure: Singling out the five first selected overlaps for $a=3$;
  the fifth (purple) shows a discontinuity. Note the vertical scale of the
  graphs. This shows that \emph{some} of the chosen eigenpairs
  (typically the lowest) are chosen correctly.\label{fig:overlaps2}}
\end{figure}

We now turn on the particle-hole interaction, by setting $f = \alpha
g$ with $\alpha>0$. In this case, the qualitative picture of the model
space overlaps shown in Fig.~\ref{fig:overlaps1} changes dramatically, 
even for very small values of 
$\alpha$. In Fig.~\ref{fig:overlaps2}, we repeat the plots from
Fig.~\ref{fig:overlaps1} for $a\leq 4$ with $\alpha = 0.05$. It is hard
to see any systematic behaviour except for the $a=2$ case. This goes
to show that even tiny changes in an operator can give big changes in
qualitative behaviour of its eigenvectors, even though the eigenvalues
are only perturbed slightly. 

In the lower right plot of Fig.~\ref{fig:overlaps2}, we have singled
out the overlaps belonging to the lowest eigenvalues for $a=3$. It
shows that there is still some ability left in the overlap selection
scheme to select some eigenvalues adiabatically. The fifth overlap
curve (and all the others not shown) has discontinuities, which would
manifest itself as discontinuities in $\hat{V}_{\mathrm{eff}}^{(3)}$. 

This trend is general, as Table \ref{tab:good-overlaps} shows. Here,
the fraction $d/m$ of the number of continuous overlaps $d$ (counted
visually) over model space dimension $m$ for different $a$ and
$\alpha$ are shown for the subspace of lowest $S=0$ or $S=1/2$ (for
$a=3$). Apparently, $a>2$ gives a smaller fraction of continuous
curves, meaning that following the eigenvalues adiabatically becomes
increasingly difficult. The fact that the two-body case exhibits such good behavior
is mainly due to the role played by the pairing force if the pairing force is stronger than the 
particle-hole contribution. With three or more active particles,
our choice of  particle-hole operator  produces strong correlations between 
the model space and the excluded space. 
\begin{table}
  \caption{Fraction $d/m$ of ``good'' selections of eigenpairs for
    $\hat{V}_{\mathrm{eff}}^{(a)}$, where $m$ is the dimension of
    $\mathcal{P}$. We limit the attention to states with quantum numbers $S=0=S_z=0$ for $a=2$,$4$,
    and $S=1/2=S_z=1/2$ for $a=3$. Also, $f=0.5g$. The $a=2$ case has a
    much larger fraction than the others, i.e., the overlap selection
    scheme works better.\label{tab:good-overlaps}}
\begin{center}
\begin{tabular}{c|ccc}
& $a=2$  & $a=3$ & $a=4$  \\ 
\hline
$g=0.05$ & $15/15$ & $4/40$ & $9/50$ \\
$g=0.1$  & $15/15$ & $4/40$ & $9/50$ \\
$g=0.5$  & $11/15$ & $4/40$ & $8/50$ \\
$g=1.0$  & $11/15$ & $4/40$ & $8/50$ \\
\end{tabular}
\end{center}
\end{table}

These considerations indicate that if any eigenvalues of
$\hat{H}_{\mathrm{eff}}^{(a)}$ can be reliable, it will only be the lowest ones. Higher
eigenvalues will almost certainly be without meaning. Of course, the
matrix elements due to the ``bad'' selections (i.e., selections after
a discontinuity in the curve) will also affect the lowest eigenvalues,
but one can hope that these effects can be neglected. From
Table~\ref{tab:good-overlaps} we thus expect $\hat{V}_{\mathrm{eff}}^{(2)}$ to perform
better than values of $a$ greater than two, but that only the lowest eigenvalues have a
reasonable error behaviour.

\begin{figure*}
\includegraphics[width=0.5\textwidth]{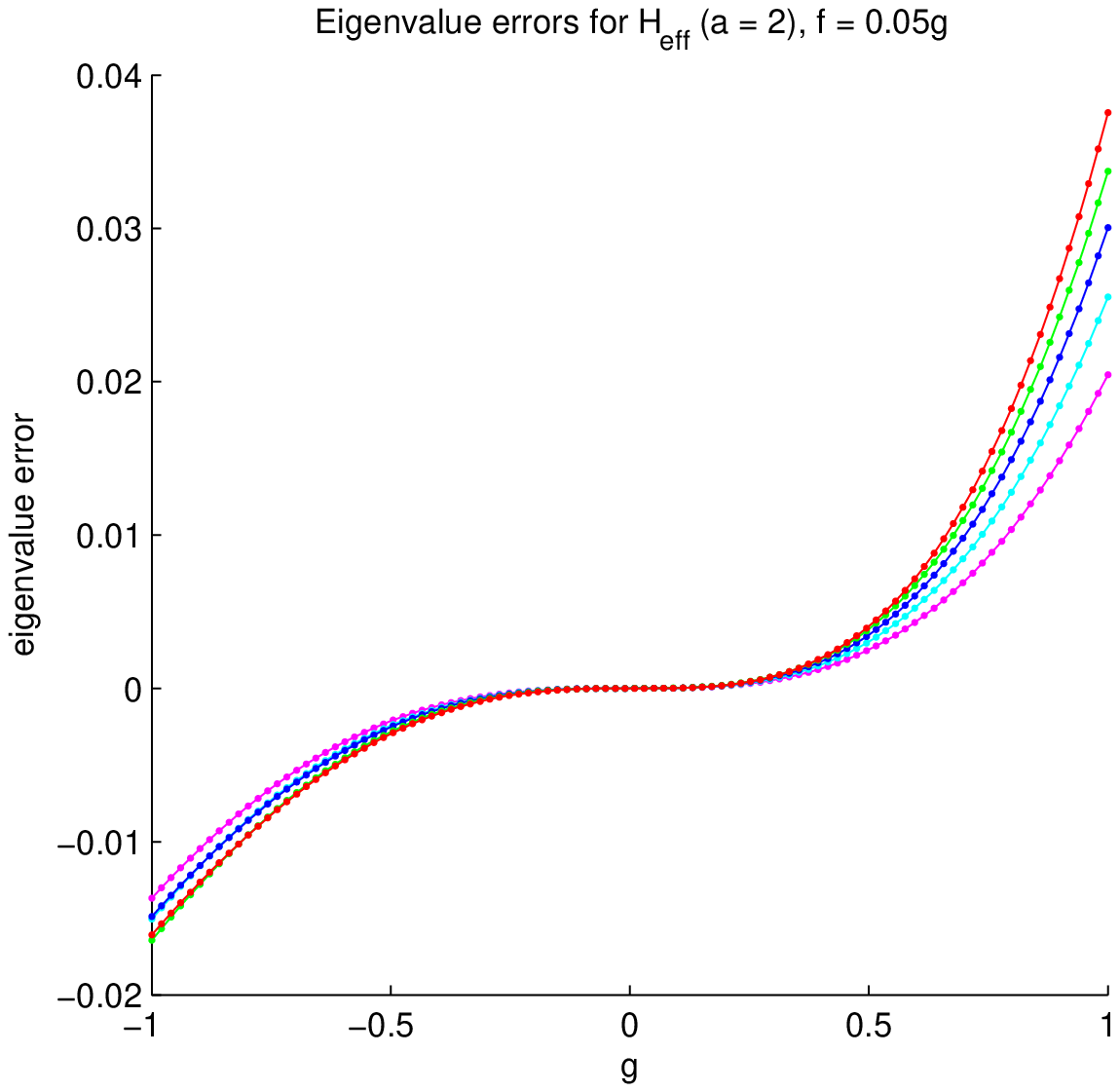}%
\includegraphics[width=0.5\textwidth]{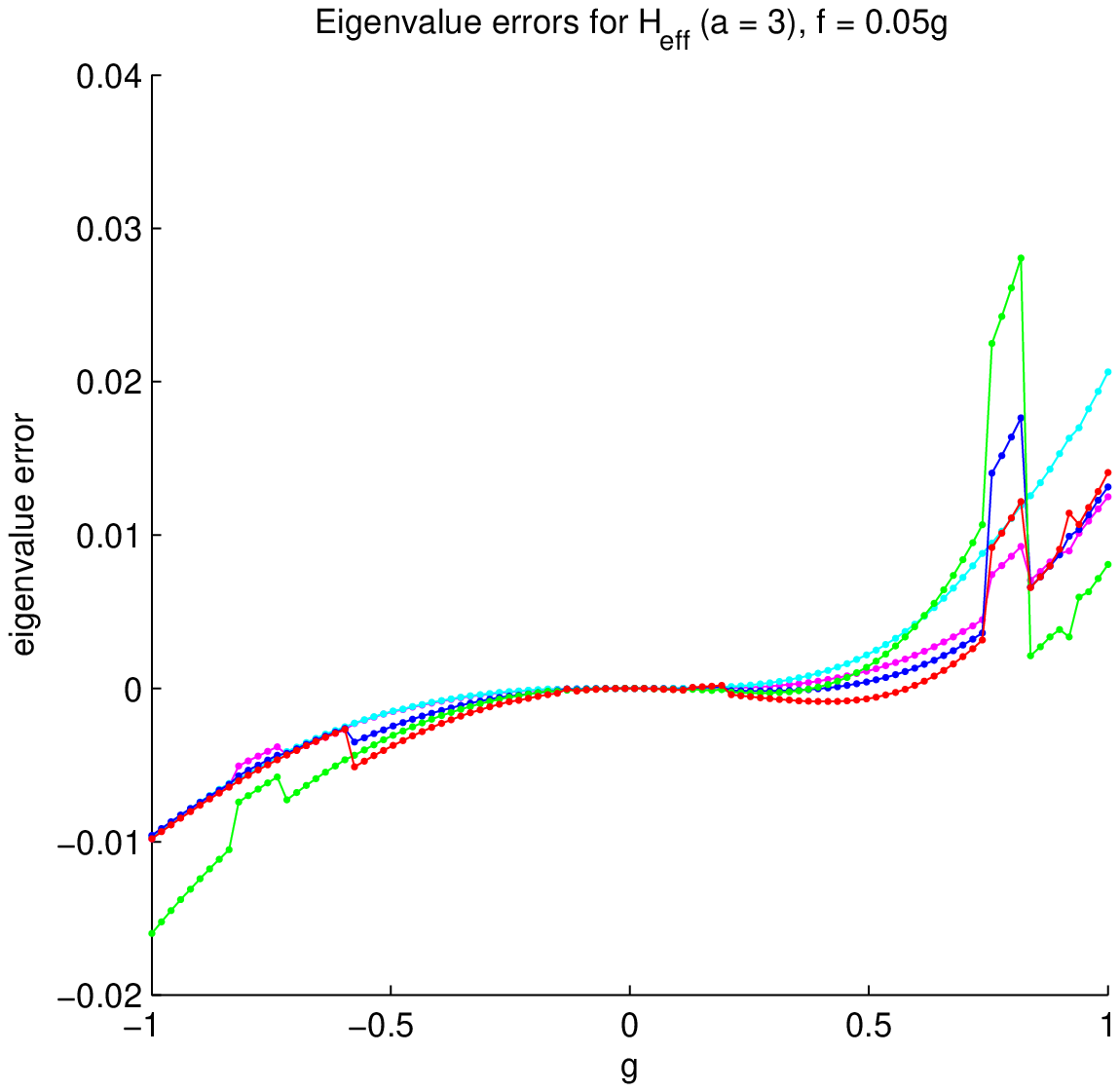}
\includegraphics[width=0.5\textwidth]{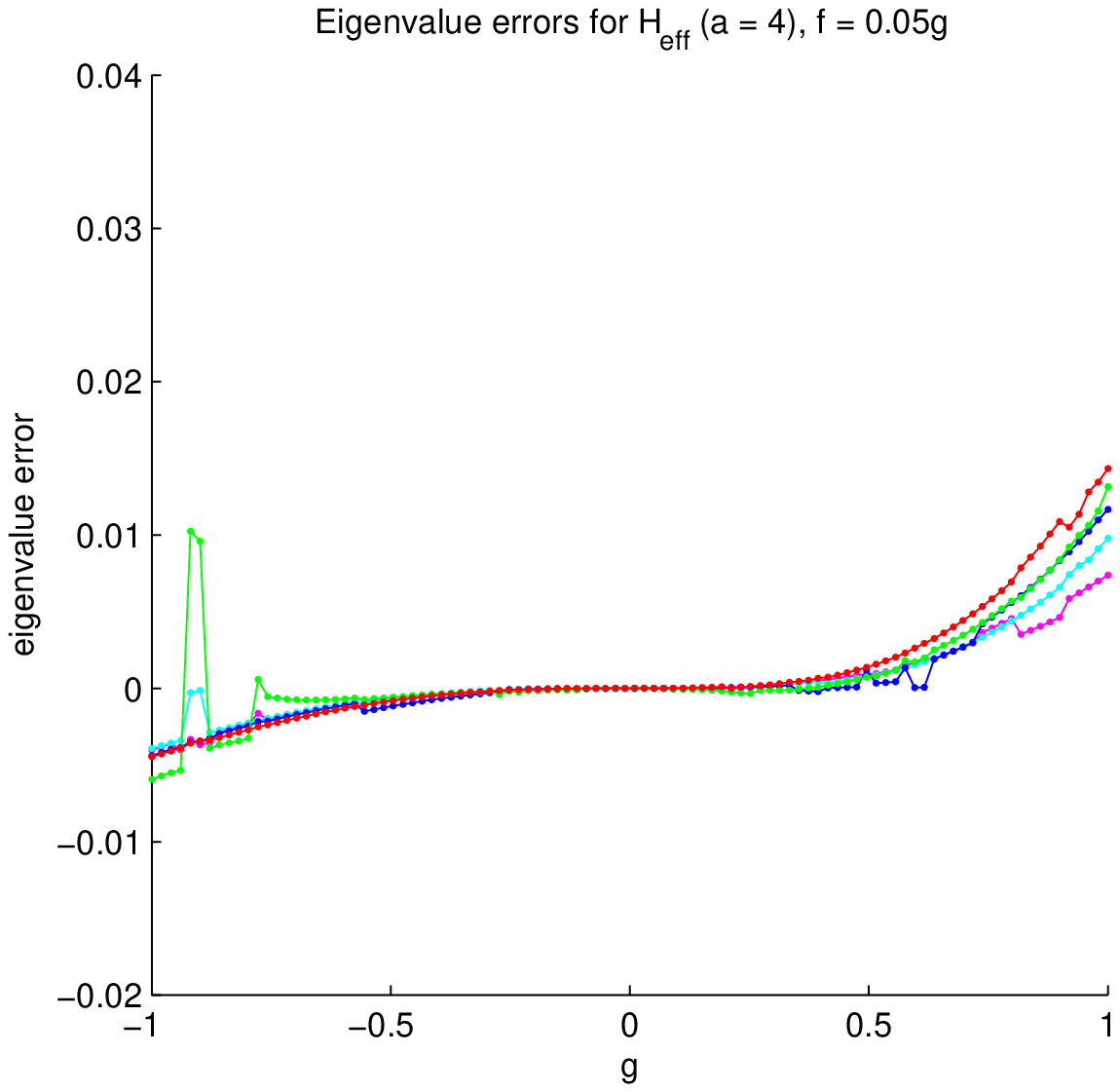}%
\includegraphics[width=0.5\textwidth]{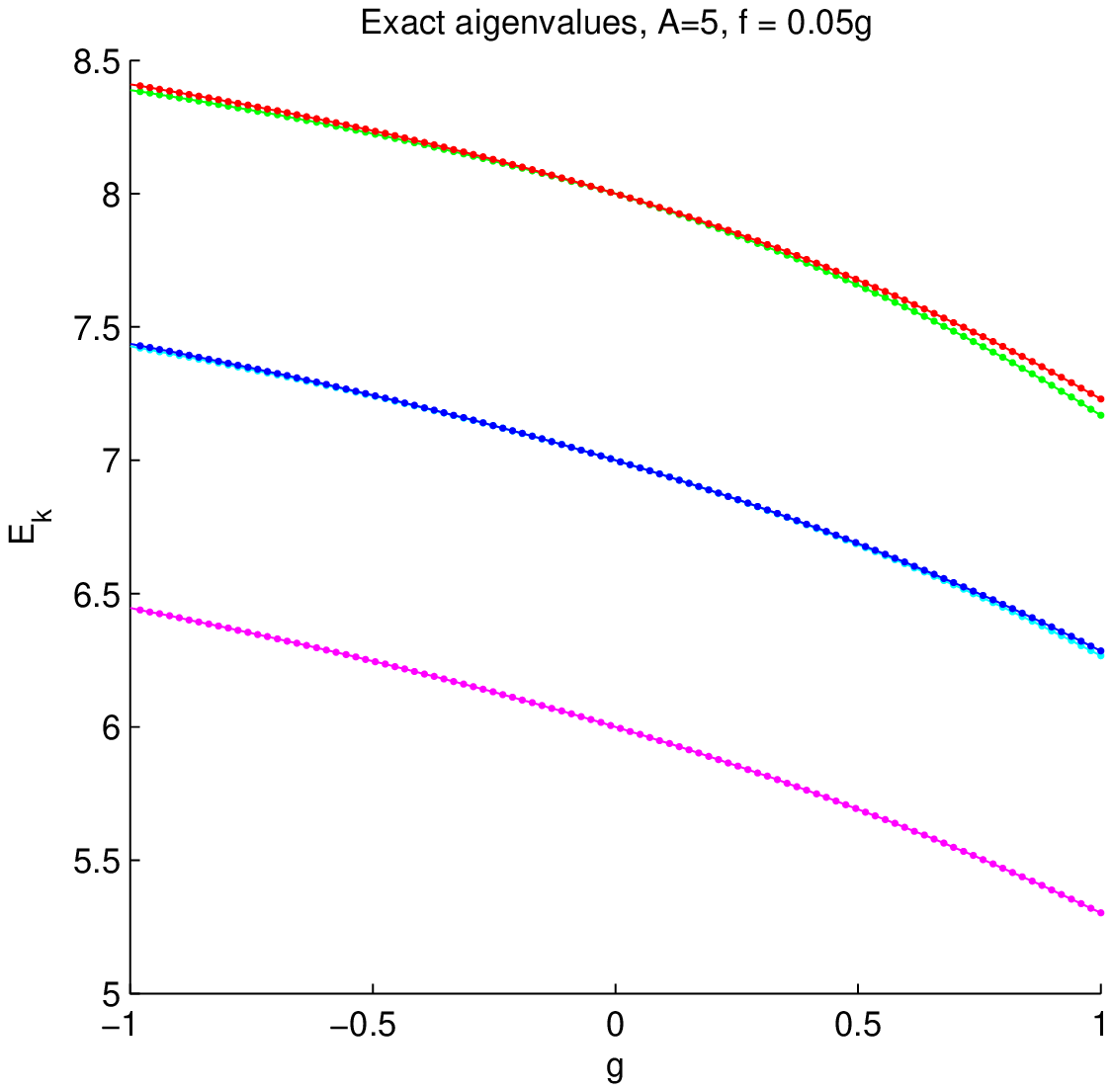}
\caption{(Color online) A few errors in eigenvalues using $\hat{V}_{\mathrm{eff}}^{(a)}$, $f=0.05g$. Lower right:
  corresponding exact eigenvalues for $A=5$. Each curve is colored
  corresponding to the exact eigenvalue curves. The $a=2$ is
  well-behaved while the  $a=3$ case and the $a=4$ case give unreliable eigenvalues.\label{fig:errors2}}
\end{figure*}

\begin{figure*}
\includegraphics[width=0.5\textwidth]{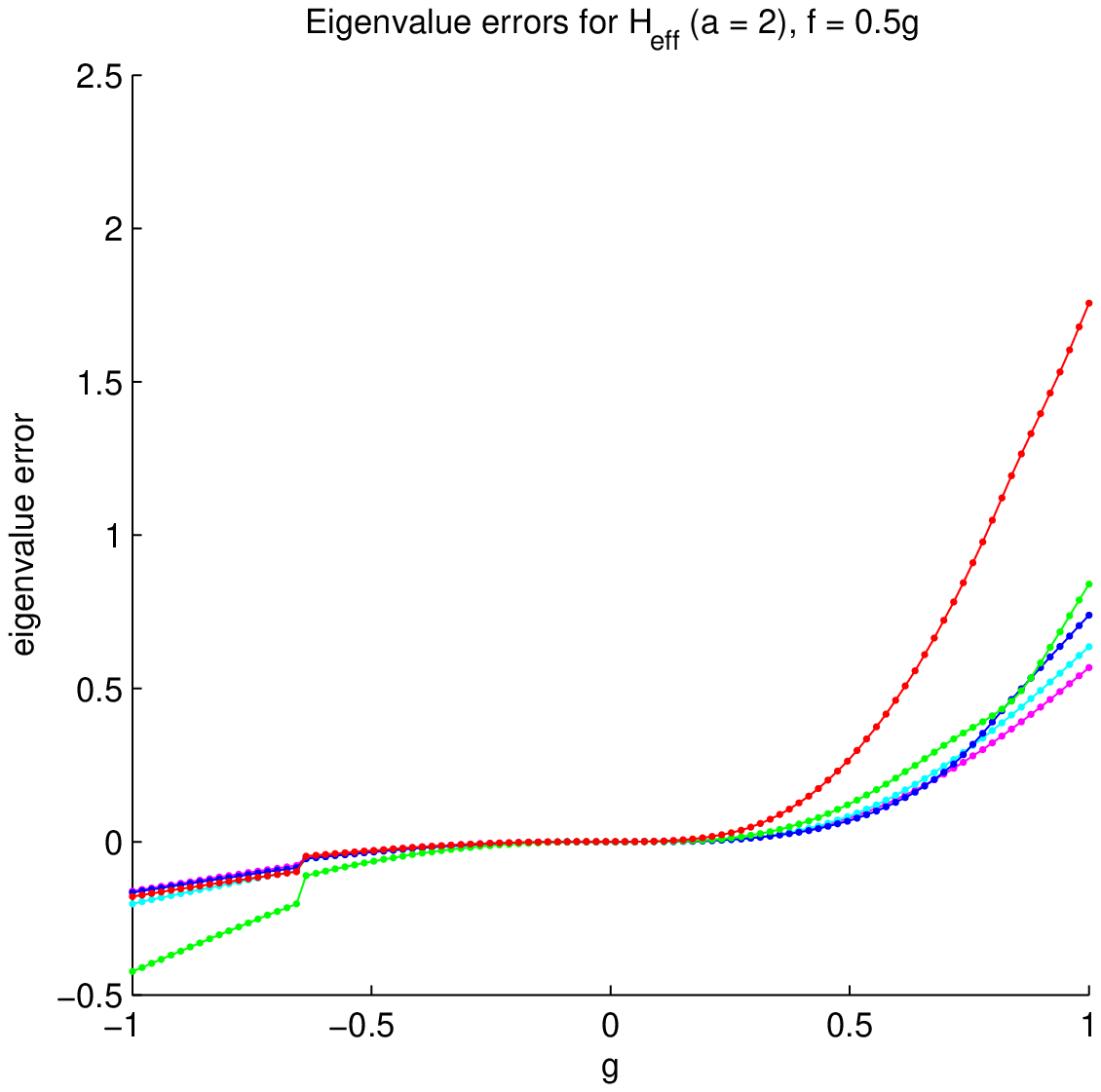}%
\includegraphics[width=0.5\textwidth]{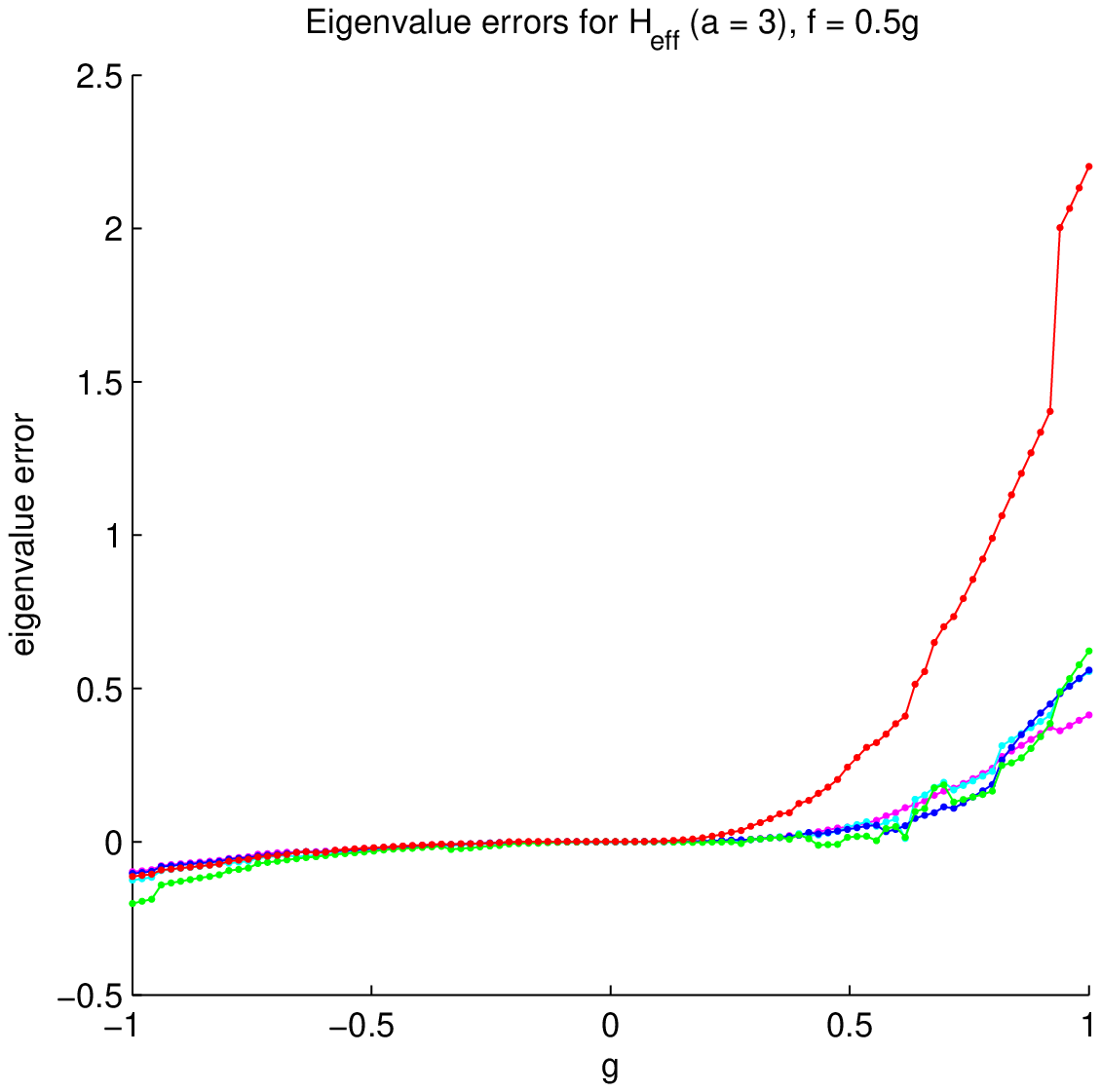}
\includegraphics[width=0.5\textwidth]{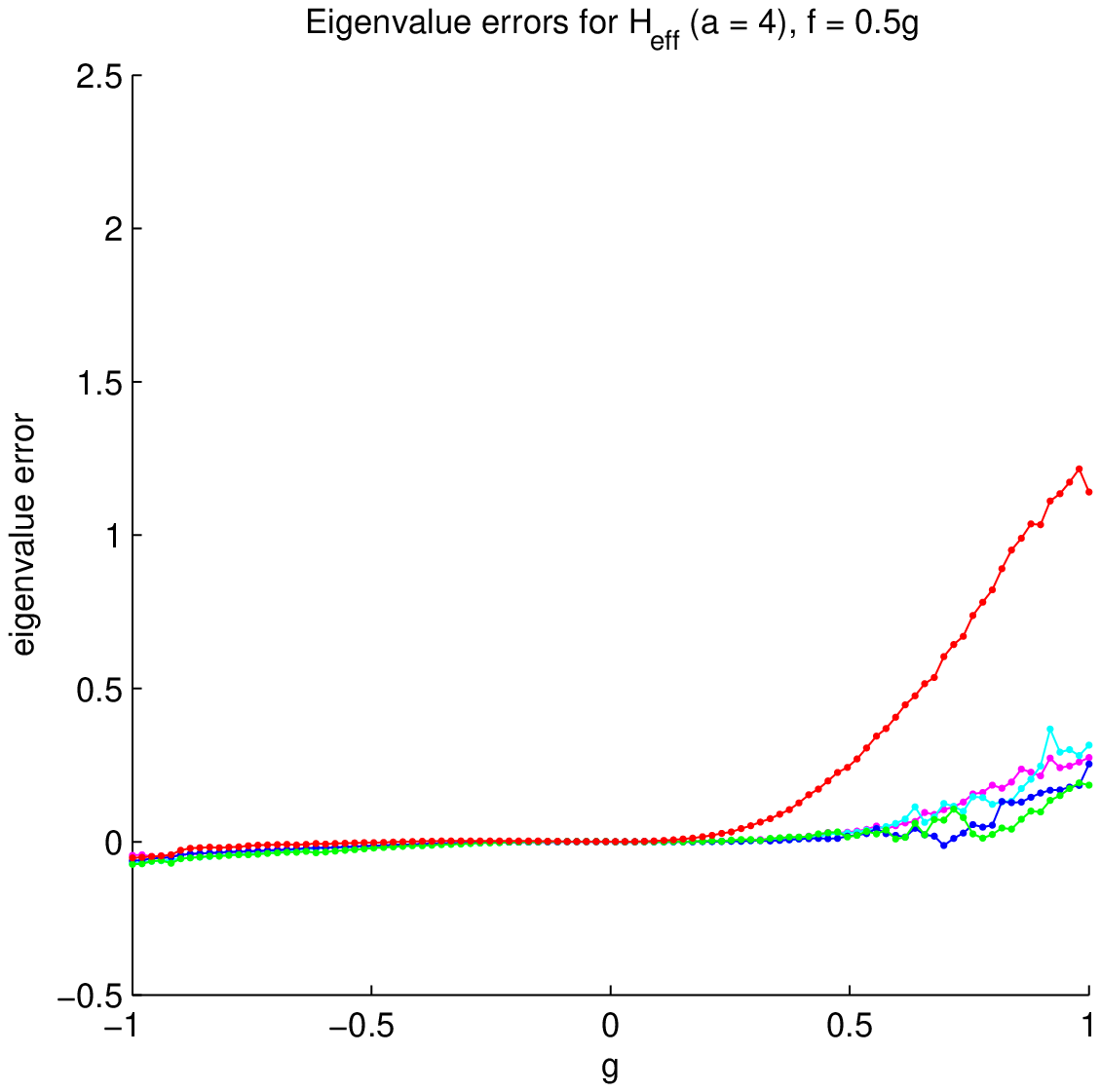}%
\includegraphics[width=0.5\textwidth]{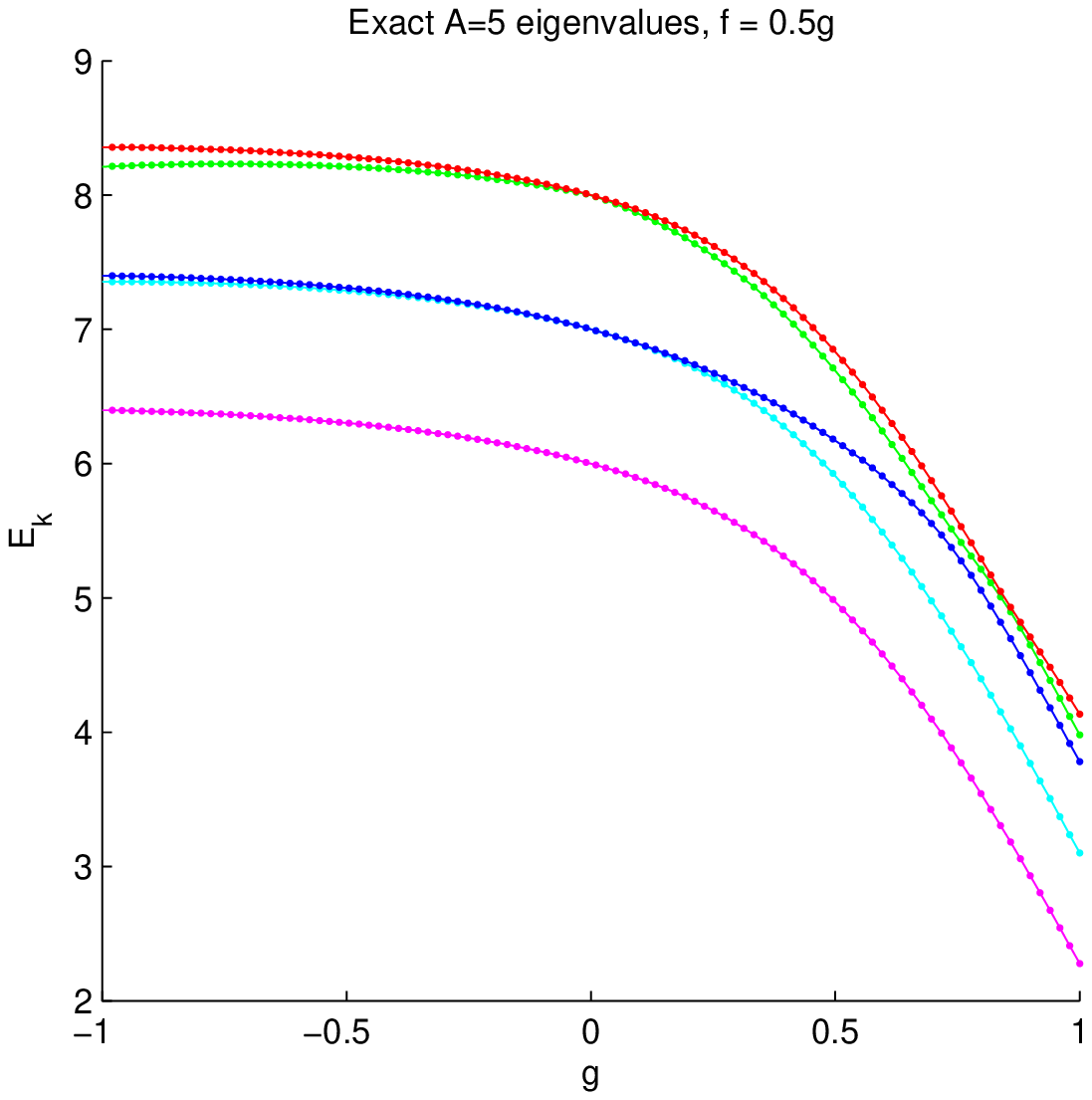}
{(Color online) A few errors in eigenvalues using $\hat{V}_{\mathrm{eff}}^{(a)}$, $f=0.5g$. Lower right:
  corresponding exact eigenvalues for $A=5$. Each curve is colored
  corresponding to the exact eigenvalue curves. The $a=2$ is now less
  well behaved. An erratic behaviour is, in all cases, amplified compared
  to the $f=0.05g$ case. Compare also the error-axis range with Fig.~\ref{fig:errors2}.\label{fig:errors3}}
\end{figure*}

In Fig.~\ref{fig:errors2} the eigenvalue error plots of Fig.~\ref{fig:errors1} are
repeated, but with $\alpha = 0.05$. Clearly, $a=2$ is able to produce
sensible results in a wide range of $g$; the other $\hat{V}_{\mathrm{eff}}^{(a)}$ have errors that jump
erratically. The picture gets worse when we try $\alpha = 0.5$, shown
in Fig.~\ref{fig:errors3}. In this case, not even the $a=2$ case will give
reliable results except for perhaps the ground state energy. We mention that
in Fig.~\ref{fig:overlaps2} the $a=2$ overlaps \emph{will}
cross at sufficiently large $\alpha$, explaining thereby the breakdown
for the $a=2$ case as well.

In summary, the particle-hole interaction changes the qualitative
behaviour of the exact eigenvectors for the $a$-body problem
drastically. It is clear that it becomes difficult to choose a model
space $\mathcal{P}$ that fits well with the model space overlaps shown in
Fig.~\ref{fig:overlaps2}, such that it becomes possible to follow the
eigenvalues adiabatically as a function of $g$. The two-body effective
interaction, however, is surprisingly well behaved for a large range of
interaction strenghts.

We have seen that using using effective interactions with $a>2$ does not seem to yield extra
accuracy in general compared to the $a=2$ case. First of all, such
$\hat{V}_{\mathrm{eff}}^{(a)}$ interactions are much harder to compute. 
Secondly, we have observed only a constant factor gain
in accuracy in the pairing case, in addition to obvious problems when the particle-hole
interaction is present.

However, with a strong particle-hole interaction, intruder-states can start playing a major role,
and effective many-body interactions with $a > 2$ are not capable of reproducing
the low-lying levels.  
The crossings of the curves in, for example, Fig.~\ref{fig:overlaps2} are due to
intruder states, and these are totally absent in integrable systems in
contrast to the apparent abundance seen here. Indeed, all two-particle
systems described by central force interactions and external harmonic
oscillator potentials are classically integrable.
There are several examples of nuclear systems where intruder states
play a major role, with perhaps the so-called {\em island of inversion} for nuclei with mass number $A\sim 31-33$ 
as one of the more popular mass regions studied recently 
(see for example the experimental results on the  
$\beta$-decay of $^{33}$Mg in Ref.~\cite{tripathi2008}). Both the parent and the daughter ($^{33}$Al) nuclei
reveal intruder-state configurations of two-particle-two-hole character in the 
lowest excited states, in addition
to proposed admixtures of one-particle-one-hole and three-particle-three-hole 
configurations for the ground state of $^{33}$Mg.  

From our results, we may therefore infer that when intruder states are present, special attention must be paid
to the construction of an effective interaction, as pointed out
 almost four decades ago by Schucan and Weidenm\"uller \cite{sw1972,sw1973}. 

To link this discussion with the topic of missing many-body correlations, 
we can view the presence of intruder states as an example of a model space which is too small.
This is in turn reflected in missing many-body correlations beyond those which can be produced by 
say a given sub-cluster effective interaction.  
Reducing the model space thus produces
missing many-body correlations, which, depending on the strength of the interaction, can produce
large differences between the exact results and  those obtained with a sub-cluster effective interaction.

We can therefore summarize by saying that an effective sub-cluster Hamiltonian will always
produce missing many-body physics. The size of these contributions can normally only be determined
{\em a posteriori}, that is after a calculation has been performed.
To be able to estimate the role of these effects is important for nuclear many-body theory, as it provides us with a
sound error estimate. 
The hope is obviously that their effect is negligible, although the calculations
of Ref.~\cite{furnstahl2009} indicate  small but non-negligible four-body contributions 
for $^4$He when a three-body Hamiltonian that reproduces the experimental binding
energy of $^3$H is employed.
When one employs nuclear forces derived from effective field theory, 
many-body terms such  as three-body interaction arise
naturally \cite{epelbaum2009a}. 
There are now clear indications from several calculations (see for example Refs.~\cite{epelbaum2009b,navratil2009,hagen2009,furnstahl2009,hagen2007,hagen2007c,pieper2001,wiringa2002}) that  at least three-body interactions have to be included in nuclear many-body calculations. Whether four-body, or more complicated many-body, interactions are important or not for an accurate description of nuclear  data is an open issue.  From a practical and computational point of view,
we obviously favour negligible four or higher many-body interactions.  

To understand how many-body interactions develop as one adds more and more particles is thus
an open and unsettled problem in nuclear many-body physics (or many-body physics in general if the 
effective Hilbert space is too small). These interactions are not observables; however, their 
role can be extracted from calculations with a  given effective sub-cluster interaction. 

To make this clear, assume that a given three-body Hamiltonian derived from effective field theory
has been used to study the chain of oxygen isotopes. With this Hamiltonian, we perform then 
precise many-body calculations for
all isotopes from $^{16}$O to $^{28}$O. 
If the errors of our calculations are negligible, we can infer that these are the results with this
specific effective Hamiltonian. 
The discrepancy between theory and experiment can then be used
to extract the role of missing many-body forces as a function of the number of nucleons.  The missing many-body physics depends obviously on the employed effective Hamiltonian.

Experiment or simple models  provide us therefore with the  benchmark our favourite theory has to reproduce. If the experimental data are reproduced to within given uncertainties, one can start analyzing the
properties of nuclei in terms of various components of the nuclear forces. This can allow for the extraction of simple physical mechanisms, as done recently by Otsuka {\em et al} \cite{taka2010}. 

This coupling between experiment and precise many-body calculations  
is what we will sketch in the next section, starting with the oxygen isotopes.

\section{Physics cases}\label{sec:physicscases}
The nuclear forces derived from effective field theory are obviously much more complicated than the above simple model. 
First of all, our simple model contains only a two-body Hamiltonian. It allowed for a numerically  exact diagonalization
for a five-particle system in a Hilbert space consisting of eight doubly degenerate single-particle states.
An effective sub-cluster interaction defined for a smaller space (typically the five lowest doubly degenerate single-particle states)
gave results which were close to the exact ones if the interaction was weak and a relevant model space was employed. 
In particular, with a particle-hole interaction,
one can easily  get a strong admixture from states outside the chosen model space. In that case, the various sub-cluster effective
interactions failed in reproducing the low-lying eigenvalues. 

The nuclear interactions derived from effective field theories
\cite{epelbaum2009a} already contain at the so-called N$^2$LO level,
three-body interactions. At the N$^3$LO level, four-body interactions appear.  These interactions are normally constructed 
for a given cutoff $\Lambda$. In principle, if one uses an interaction at the  N$^3$LO level, one should also
include four-body interactions. Normally, only three-body interactions at the N$^2$LO level are introduced, together with a 
two-body interaction that fits two-particle data (scattering data and bound state properties) at the N$^3$LO level. 

This means that when employed in a many-body context with more than three particles, our Hamiltonian lacks some many-body correlations.
Hopefully these are small. A comparison with data, if our theory produces converged results at a given level of many-body physics, 
may then reveal their importance.  

In this sense, a given nuclear interaction computed with effective field theory at a given N$^n$LO level, should contain all possible
many-body interactions. At the N$^3$LO level, one should include one-body, two-body, three-body, and four-body interactions. 
Omitting some of these leads to a sub-cluster effective Hamiltonian.  In a certain sense, this parallels the discussion from the previous  section. However, in the nuclear physics case, it is important to keep in mind that these interactions
have been derived with a specific energy cutoff.  The energy cutoff defines our effective Hilbert space, or model space.  
If the model space defined for an actual many-body calculation does not include all possible excitations within this model
space, one can easily end up with the type of missing many-body correlations  discussed in the previous section. 

To make our scheme more explicit, 
assume that we are planning a calculation of  the oxygen isotopes with an effective field theory interaction 
that employs a cutoff $\Lambda=600$ MeV.  Assume also that our  favourite single-particle basis is the harmonic oscillator, with
single-particle energies $\varepsilon_{nl}=\hbar\omega(2n+l+3/2)$, with  $\omega$ the oscillator frequency, $n=0,1,2,\dots$ being the 
number of nodes and $l$ the single-particle orbital momentum.  The oscillator length $b$ is defined as
\[
b=\sqrt{\frac{\hbar}{m\omega}}.
\]
We define $p=0,1,2,\dots,P$ with $P=2n+l$ as the quantum number 
$p$ of the highest-filled level.  The level labelled $p$ can accommodate $(p+1)(p+2)$ fermions, with a spin degeneracy of two
for  every single-particle state taken into account.
For a given maximum value of $P=2n_{\mathrm{max}}+l_{\mathrm{max}}$, we have a total of 
\[
N=\sum_{p=0}^P(p+1)(p+2)= \frac{(P + 1)(P + 2)(P + 3)}{3},
\]
single-particle states. 

The cutoff $\Lambda$ defines the maximum excitation energy  a system of $A$ nucleons can have. The largest single-particle
excitation  energy (corresponding to possible one-particle-one-hole correlations) is then 
\[
\Lambda = \hbar\omega P= \hbar\omega (2n_{\mathrm{max}}+l_{\mathrm{max}}).
\]
The value of $\hbar\omega$ can be extracted from the mean-squared radius of a given nucleus.  One can show that this results in \cite{kirson2007}
\[
\hbar \omega \approx  \left(\frac{3}{2}\right)^{4/3}\frac{\hbar^2}{2m_Nr_0^2}A^{-1/3},
\] 
with $r_0\approx 1$ fm. Setting $A=16$ and $\Lambda=600$ MeV results in 
$P\approx 42$.  The largest possible value for $n$ is then $n_{\mathrm{max}}\approx 21$, or 22 major shells.
With $P=42$, the total number of single-particle states in this model space is $28380$!
  
For $^{16}$O, this means that we have to distribute eight protons and eight neutrons in $28380$ single-particle states, respectively.
The total number of Slater determinants, with no restrictions on energy excitations, is
\[
\left(\begin{array}{c}28380\\8\end{array}\right)\times \left(\begin{array}{c}28380\\8\end{array}\right)\approx 10^{62}.
\]
Any direct diagonalization method in such a huge basis is simply impossible. One possible approach is to introduce  a 
smaller model space with 
a pertinent effective 
interaction,  similar to what was discussed in the previous section. Depending on the size of the model space and the strength 
of the interaction, this can 
lead to uncontrolled many-body correlations, with effects similar to what was discussed within the framework of the simple model
from Section \ref{sec:toymodel}.
This has been the philosophy of the no-core shell-model approach \cite{navratil2009} or many-body perturbation theory \cite{hko1995}.
Another alternative is to employ methods which allow for systematic inclusions to all orders of specific sub-clusters of correlations
within the full model space defined by $\Lambda$. This reduces the computational effort considerably and accounts, at the same time,
for most of the relevant degrees of freedom. The Coupled cluster method \cite{bartlett2007,dean2004,hagen2008} 
and Green's function theory \cite{barbieri2009} allow for such systematic 
expansions starting from a given Hamiltonian.

Coupled-cluster theory has been particularly successful in both quantum chemistry and nuclear physics. It fulfills basically all the requirements we listed in Section \ref{sec:introduction}. Furthermore, with a given truncated 
single-particle basis (our effective Hilbert space) we can extract precise error estimates on the energy \cite{kvaal2009,schroeder2008}.
It is a topic of current research to be able to predict and understand the error due to truncations in the number of many-body correlations.
As an example, most coupled-cluster calculations seldom go beyond correlations of the three-particle-three-hole type, the so-called
triples correlations.  An estimate of the error made due to the omission of four-particle-four-hole correlations (and higher-order correlations) would 
therefore be very useful. That would lead to a much more  predictive theory. 

\subsection{Brief review of coupled cluster theory}

Our many-body method of choice is thus coupled-cluster theory, see Refs.~\cite{bartlett2007,dean2004,hagen2008} for more details.
The coupled-cluster
method fulfills Goldstone's linked-cluster theorem and therefore
yields size-extensive results, i.e., the error due to the truncation is
linear in the mass number $A$ of the nucleus under
consideration. Size extensivity is an important issue when approximate
solutions to all but the lightest nuclei are
sought~\cite{bartlett2007,dean2008}. Second, the computational effort scales
gently (i.e., polynomial) with increasing dimension of the
single-particle basis and the mass number $A$. The method has met
benchmarks in light nuclei~\cite{hagen2007,mihaila2000}. 

Our algorithm is as follows:
\begin{enumerate}
\item We start with a nuclear force from effective field theory at a given N$^n$LO order. In the calculations we present here, we 
include only a nucleon-nucleon
interaction to order N$^3$LO. 
We neglect
three-nucleon forces since their application within the
coupled-cluster method is still limited to smaller model
spaces~\cite{hagen2007c}.   This gives rise to one source of systematic error and most likely the largest uncertainty in our results.
So-called power-counting estimates
from chiral effective field theory  result in an uncertainty of about 2~MeV per nucleon, see Ref.~\cite{hagen2009} for further details.
\item We compute ground state properties and excited states by including one-particle-one-hole, two-particle-two-hole, and selected 
three-particle-three-hole correlations. These correlations are called singles, doubles, and triples, respectively. 
We consider 
corrections due to triples excitations within
the so-called $\Lambda$CCSD(T) approximation, see for example \cite{hagen2009} for more details. Comparison of a calculation which includes only one-particle-one-hole and two-particle-two-hole correlations (CCSD), see Ref.~\cite{hagen2009}, shows that 
triples corrections account for 10-15\%. Similar ratios are
found in coupled-cluster calculations of atoms and
molecules, and experience in quantum chemistry (see for example
Ref.~\cite{bartlett2007}) suggests that the truncation of the cluster
amplitudes beyond the triples corrections introduces an error of a few
percent. 
\item The actual single-particle space used  is defined close to the effective field theory model space.  Our error here is smaller than the
error made in the previous two points. The final results are more or
less converged as a function of the size of the model space.
A precise error estimate for a given single-particle basis was presented within a configuration interaction approach by Kvaal \cite{kvaal2009}. Similar error estimates apply to coupled-cluster theory as well \cite{schroeder2008}.
\end{enumerate}

Within the choice of model space, most of our expectation values are practically converged. This is, for example, reflected in the fact
that the ground state energy is almost independent of the chosen harmonic oscillator energy (for demonstrations, see for example
Refs.~\cite{hagen2009,hagen2007,hagen2008}). 
From this, we can infer that our specific many-body approximation yields probably the optimal result which can be achieved at this specific level of many-body physics.
Eventual discrepancies with experiment can then most likely be attributed to missing many-body correlations. A typical case would be the 
lack of three-body interactions  from effective field theory.

\subsection{Oxygen isotopes}
The oxygen isotopes are the heaviest isotopes for 
which the drip line is well established. There are large experimental 
campaigns worldwide \cite{elekes2007,schiller2006,hoffman2008} which
aim at uncovering the properties of the oxygen isotopes, both at or
close to the drip line and beyond. Two out of four stable even-even
isotopes exhibit a doubly magic nature, namely $^{22}$O ($Z$=$8$,$N$=$14$) 
\cite{thirolf2000} and $^{24}$O ($Z$=$8$, $N$=$16$) \cite{stanoiu2004,hoffman2009}.   
The structure of these two doubly magic nuclei is assumed to be governed 
by the evolution of the $1s_{1/2}$ and $0d_{5/2}$  one-quasiparticle states.  
The isotopes $^{25-28}$O 
are all
believed to be unstable towards neutron emission, even though $^{28}$O
is a doubly magic nucleus within the standard shell-model picture. 

\begin{figure}[h]
\begin{center}
\includegraphics[width=0.5\textwidth,clip=]{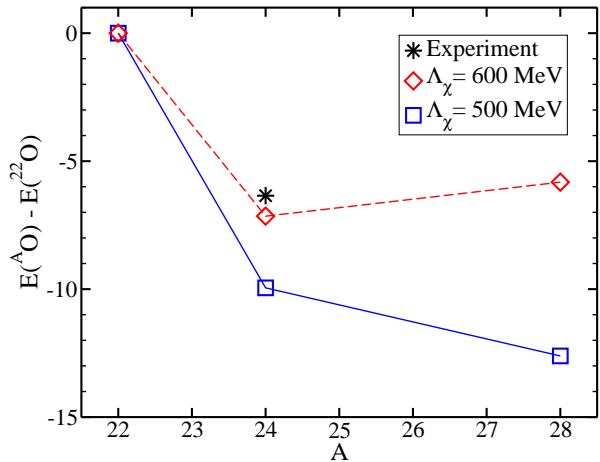}
\caption{(Color online) Ground-state energies of neutron-rich oxygen
isotopes $^A$O relative to $^{22}$O for chiral
interactions with two different cutoffs $\Lambda_\chi$. Taken from Ref.~\cite{hagen2009}.}
\label{fig:oxisotopes}
\end{center}
\end{figure}

Of interest to us is the fact that we can perform {\em ab initio} 
coupled-cluster calculations for all assumed closed-shell nuclei
of this isotopic chain, that is, oxygen isotopes with $A=16$, $A=22$, $A=24$, and $A=28$. Furthermore, we can also 
compute the $A\pm 1$ nuclei such as $^{25}$F, $^{25}$O, and  $^{23}$N. 
This means further that we can extract the isospin dependence of say spin-orbit partners. 
In Ref.~\cite{hagen2009} we considered the nuclei $^{16,22,24,28}$O and computed
their ground-state energies within the $\Lambda$CCSD(T) approximation
for chiral interactions with cutoffs of $\Lambda_\chi=500$~MeV and
$\Lambda_\chi=600$~MeV, respectively. Figure \ref{fig:oxisotopes} shows the ground state energies relative to
$^{22}$O.
At a cutoff $\Lambda_\chi=500$~MeV, we find that $^{28}$O is bound by
about 2.7~MeV with respect to $^{24}$O.  However, the situation is
reversed at the higher cutoff $\Lambda_\chi=600$~MeV, and the
difference is about -1.3~MeV. With the uncertainties due to missing three-body interactions, 
it is presently
not possible to reach a conclusion regarding the existence of
$^{28}$O.  However, using interactions from chiral effective field theory,  we see that the
stability of $^{28}$O depends mainly on the contributions of the
three-nucleon force (and probably more complicated many-body forces), and that even small contributions can tip the
balance in either direction. A mass measurement of $^{28}$O is clearly needed. 

The above results show that we do not yet understand fully how many-body forces evolve as we add more nucleons. 
The hope is that the inclusion of a three-body interaction from effective field theory reduces the dependence on
the cutoff at the two-body level. This was nicely demonstrated by Jurgenson {\em et al} \cite{furnstahl2009} in a recent study 
of $^3$H and $^4$He. For $^4$He, there were indications that a small four-body interaction may be needed, as also demonstrated by
Epelbaum {\em et al} \cite{epelbaum2009b}. 
Our {\em ab initio} calculations indicate also that  the recent results
from phenomenological shell-model approaches regarding the unbound
character of $^{28}$O should be viewed with caution. The combination
of three-nucleon forces, the proximity of the continuum, and the
isospin dependence are a challenge for reliable theoretical predictions, see also Refs.~\cite{taka2009,koshiroh2009}.

Summarizing, we can claim that there is {\em a  posteriori} evidence
of the need of a three-body interaction or more complicated
interactions.
We hope that with the inclusion of three-body forces, the difference seen for different chiral interactions should become small.

Furthermore, if the differences between theory and experiment are small, theory can be used to extract simple information on say
various components of the nuclear force.
We can, for example, study the evolution of many-body forces as we increase the 
number of valence nucleons. One possibility is to extract  the $A$-dependence and the isospin dependence 
of such correlations as we move towards the drip line. Another interesting quantity which can be extracted is 
the isospin and $A$-dependence of the spin-orbit force.
This will allow us to answer and understand which mechanisms change
the single-particle fields close to the drip line.
For these mass regions and the oxygen isotopes, experimental data on $^{28}$O are very important. Similarly, for spin-orbit partners,
measurements of closed-shell nuclei with one particle added or removed beyond $A=24$ will allow us to benchmark theory
with experiment and perhaps understand which parts of the nuclear
force are important close to the drip line.

\subsection{Calcium and nickel isotopes}

Doubly magic nuclei are particularly important
and closed-shell nuclei like $^{56}$Ni, $^{100}$Sn and $^{132}$Sn have been the focus of several experiments
during the last several years~\cite{ni56a,Yur.06,ni56c,ni56d,ni56e,ni56f,sn100a,sn100b,sn100c}. Their structure provides  
important information on theoretical interpretations  and our basic
understanding of matter.  
In particular, recent experiments~\cite{ni56a,Yur.06,ni56c,ni56d,ni56e,ni56f} have aimed at extracting 
information about single-particle degrees of freedom in the vicinity of $^{56}$Ni. Experimental information
from single-nucleon transfer reactions  
and magnetic moments \cite{ni56a,Yur.06,ni56c,ni56d,ni56f} can be used to extract and interpret complicated many-body wave 
functions in terms of effective single-particle degrees of freedom. Transfer reactions provide, for example,
information about the angular distributions, the excitation energies, and the spectroscopic factors of possible single-particles states.

Much of the philosophy exposed in the previous subsection can be repeated for studies of calcium and nickel isotopes.
The interesting feature here is that we probe systems with more nucleons. The chain of nickel isotopes
exhibits four possible closed-shell nuclei, namely $^{48}$Ni, $^{56}$Ni, $^{68}$Ni,
and  $^{78}$Ni, with a difference of 30 neutrons from  $^{48}$Ni to $^{78}$Ni. 
Similarly, for calcium isotopes in the $fp$-shell, 
we have 20 neutrons that can be added to $^{40}$Ca and five possible closed-shell nuclei,
$^{40}$Ca, $^{48}$Ca, $^{52}$Ca, $^{54}$Ca and $^{60}$Ca. The binding energies of $^{54}$Ca and $^{60}$Ca are 
presently not available.
Moreover, compared with the oxygen isotopes with at most 12 nucleons in the $sd$-shell outside $^{16}$O, 
different single-particle degrees of freedom are probed.  Comparing theory and experiment can again
provide important information about spin-orbit partners close to the
drip line, with information on 
their density and isospin dependence as well.

As stated in the previous subsection, we are in the position where such nuclei can be calculated rather
accurately with a given two-body Hamiltonian. This is demonstrated in Refs.~\cite{barbieri2009,hagen2008}. 
We are not limited to closed-shell systems but can also compute ground states and excited states of $A\pm 1$ ($A$ is the closed-shell nucleus) nucleons rather precisely. It means that a nucleus like $^{79}$Cu can be accessed with {\em ab initio} methods like 
coupled cluster theory.  Single-particle properties like magnetic moments and 
spectroscopic factors can then give a measure of how good a closed
nucleus $^{78}$Ni is.
Experimental data which probe single-particle properties on $A\pm 1$ systems close to the dripline, will therefore 
provide important benchmarks for theory.

\subsection{Ab initio studies of nuclei around mass $A=100$}

We end this section by stating that the analysis which can be performed for $fp$-shell and $sd$-shell nuclei
can be extended to the region of the tin isotopes as well, 
with both $^{100}$Sn and $^{132}$Sn as two very important nuclei
for our understanding of the stability of matter, see for example the recent works of Refs.~\cite{sn100a,sn100b,sn100c,sn132}. 
In this case we expect to be able to run similar calculations
within the next two to three years for nuclei  
like $^{100}$Sn, $^{132}$Sn, and $^{140}$Sn together with   
the corresponding $A\pm 1$ nuclei.  We can then test the development of many-body forces for an even larger chain of isotopes
and provide theoretical benchmarks for nuclei near $^{140}$Sn, of great importance for the understanding of the $r$-process, a nucleosynthetic process responsible
for the production of around half of the heavy elements. A recent experiment on $^{132}$Sn
shows clear evidence that this nucleus is an extremely good closed-shell nucleus \cite{sn132}.  This has, in turn, consequences for 
our understanding of quasi-particle properties in this mass region.
We mention also here that  
$^{137}$Sn is the last reported neutron-rich isotope (with half-life). Our final aim is to provide reliable predictions for all possibly closed-shell nuclei, from $A=4$ to $A=208$. 

\section{Conclusions}

The aim of this article has been to shed light on our understanding of many-body correlations in nuclei. 
Since all theoretical calculations
involve effective Hamiltonians and effective Hilbert spaces, it is crucial to have a handle on the role many-body correlations play in a many-body system. 
This means that a sound
theory should provide error estimates on the importance of neglected many-body effects. To understand these and develop  sound error estimates is mandatory if one wants to have a predictive theory.

We are now in a position where fairly precise results can be obtained for several closed-shell nuclei with a given two-body Hamiltonian.  
However, since most two-body interactions used nowadays are based on chiral effective field theory, it means that three-body and,
more complicated, many-body interactions arise at different chiral orders \cite{epelbaum2009a}. The neglect of such higher-order terms
leads to different predictions for two-body Hamiltonians. Experiment can then be used to understand how important these 
neglected degrees of freedom are.  Stated in a more philosophical way,
we seem to be doing pretty well at 
`doing the problem right' (verification); but we still struggle with
`validation' (doing the right problem). Having now developed the tool
set to rigorously solve the nuclear many-body problem accurately, we can
use these tools to more fully investigate the nuclear
interaction.

The hope is that the addition of three-body interactions will produce results which are more or less independent of the cutoff
used in chiral effective field theory. This means, in turn that if our theoretical calculations with a three-body Hamiltonian 
reproduce, for example, experimental
binding energies to high accuracy (theoretical and compared with experiment), we can then start analyzing properties like 
single-particle energies close to the drip line or the $A$-dependence of missing many-body correlations.  
Our formalism allows us now to compute closed-shell nuclei along an isotopic chain
and extract single-particle energies and properties. For the chains of oxygen, calcium, nickel, and tin isotopes, we can therefore  provide
important information about nuclear structure at the drip lines of these chains. 
But in order to do this, one needs experimental data on properties like
the binding
energies of $^{28}$O or $^{54}$Ca. We feel that the enormous progress
which has taken place in the last few years in nuclear theory can really lead to a predictive and reliable approach  to nuclear many-body systems. In this endeavor, the close ties between  theory and experiment
are crucial in order to understand properly the limits of stability of matter.

\section*{Acknowledgments} We are much indebted to Thomas Papenbrock
for many discussion and deep insights on nuclear many-body
theories. Research sponsored by the Office of Nuclear Physics,
U.S.~Department of Energy and the Research council of Norway.

\end{document}